\documentclass[11pt]{article}
\usepackage[utf8]{inputenc}
\usepackage{amsmath,amssymb,mathrsfs}
\usepackage{geometry}
\usepackage{xcolor}
\usepackage{tikz}
\usepackage{amsmath}
\usepackage{nicematrix}
\usepackage{mathtools} 
\usepackage{comment}
\usepackage{cite}
\usepackage[dvipsnames]{xcolor}
\definecolor{myblue}{RGB}{30,80,200}
\usepackage[
    colorlinks=true,
    linkcolor=myblue,
    citecolor=myblue,
    urlcolor=myblue
]{hyperref}

\numberwithin{equation}{section}

\newcommand{\be}{\begin{equation}}
\newcommand{\ee}{\end{equation}}
\newcommand{\bea}{\begin{eqnarray}}
\newcommand{\eea}{\end{eqnarray}}

\renewcommand{\arraystretch}{1.2}
\newcommand{\calL}{\mathcal{L}}

\newcommand{\bbR}{\mathbb{R}}

\definecolor{colorHS}{rgb}{0.9,0,0}

\begin{document}

\begin{titlepage}

$ $
\vspace{3mm}

\begin{center}

    {\LARGE \sc Minimal Massive Gravity  \\[4mm]{}coupled to higher spins}\\[10mm]

{\large

Franz Ciceri${}^{a}$, Nihat Sadik Deger${}^{b,c}$, Bastien Duboeuf${}^{d}$,\\ and Henning Samtleben${}^{e,f}$}

\vspace{10mm}

${}^a${\it Laboratoire de Physique Théorique et Hautes Energies, Sorbonne Université \\  F-75005 Paris, France}

\vskip 2 ex

${}^b${\it Department of Mathematics, Bogazici University \\ Bebek, 34342, Istanbul, T\"urkiye}

\vskip 2 ex

${}^c${\it Feza Gursey Center for Physics and Mathematics, Bogazici University \\ Kandilli, 34684,
Istanbul, Türkiye}

\vskip 2 ex

${}^d${\it Max-Planck-Institut f\"ur Gravitationsphysik \\ Am M\"uhlenberg 1, DE-14476 Potsdam, Germany}

\vskip 2 ex

${}^e${\it ENSL, CNRS, Laboratoire de physique, F-69342 Lyon, France}

\vskip 2 ex

${}^f${\it Institut Universitaire de France (IUF)}

\end{center}

\vskip10ex
\begin{center}
\begin{minipage}{0.85\textwidth}
\small
Among three-dimensional massive gravities, Minimal Massive Gravity (MMG) is distinguished by a parameter regime in which the AdS bulk theory and the dual conformal field theory can be simultaneously unitary. We couple MMG to a finite tower of higher-spins by promoting the fields of its first-order formulation to $\mathfrak{sl}(N)$-valued ones. We also present an extended formulation involving a group-valued St\"uckelberg scalar, which realizes the higher-spin translations broken by the massive deformation. We then construct AdS$_2\times$S$^1$ solutions carrying higher-spin hair specific to the massive theory. Finally, we analyse the mass spectrum of the higher-spin modes around AdS$_3$.
\end{minipage}
\end{center}

\begingroup
\renewcommand\thefootnote{}
\long\def\@makefntext#1{#1}
\footnotetext{%
\hspace*{-1.8em}\texttt{ciceri@lpthe.jussieu.fr}\\
\texttt{sadik.deger@bogazici.edu.tr}\\
\texttt{bastien.duboeuf@aei.mpg.de}\\
\texttt{henning.samtleben@ens-lyon.fr}}
\addtocounter{footnote}{-1}
\endgroup

\end{titlepage}


\tableofcontents \noindent {}

\section{Introduction}
In three dimensions (3D) the Weyl tensor vanishes identically, so that pure Einstein gravity has no local degrees of freedom and every solution of the vacuum field equations with a cosmological constant is locally maximally symmetric \cite{Deser:1983tn, Deser:1983nh}. What remains is a theory of global and boundary degrees of freedom, simple enough to admit an exact treatment \cite{Witten:1988hc} yet rich enough to retain characteristic problems of quantum gravity \cite{Carlip:1998uc}. For a negative cosmological constant the theory admits the BTZ black hole \cite{Banados:1992wn}, and the asymptotic symmetries of AdS$_3$ form two copies of the Virasoro algebra \cite{Brown:1986nw}, an observation that anticipated the AdS/CFT correspondence \cite{Maldacena:1997re}. Holography is particularly powerful in this setting, since the dual conformal field theory (CFT) is two-dimensional and its infinite-dimensional symmetry algebra strongly constrains the dynamics.

Three dimensions is also special from the higher-spin perspective. In dimensions $D\geq 4$, all known interacting theories of massless higher-spin fields require infinite towers of fields, whether in (anti-)de Sitter space, as in Vasiliev's equations (for reviews see \cite{Sezgin:2002rt, Sorokin:2004ie, Vasiliev:2004cp, Bekaert:2004qos, Bekaert:2010hw}), or in flat space, as in chiral higher-spin gravity \cite{Ponomarev:2016lrm}.
 In 3D, by contrast, massless fields of spin $s\geq 2$ carry no local degrees of freedom, and consistent interacting theories with a finite number of higher-spin fields exist \cite{Aragone:1983sz, Blencowe:1988gj}. Einstein gravity with a negative cosmological constant can be formulated as a Chern-Simons theory with gauge group SL(2)$\times$SL(2) \cite{Achucarro:1987vz, Witten:1988hc}, and replacing this group by SL($N$)$\times$SL($N$) with $N>2$ yields, for the principal embedding of $\mathfrak{sl}(2)$ into $\mathfrak{sl}(N)$, gravity coupled to symmetric tensor fields of spins $3,\dots,N$ \cite{Campoleoni:2010zq}. Properties of this model were studied further in \cite{Campoleoni:2010zq,Campoleoni:2011hg,Ammon:2011nk,Campoleoni:2012hp,Tan:2011tj}.

A propagating graviton can be introduced by adding higher-derivative terms to the Einstein-Hilbert action. Such theories, however, generically propagate ghosts around their maximally symmetric vacua, since their field equations contain third- or higher-order time derivatives \cite{Stelle:1976gc, Stelle:1977ry}. A few exceptions are known in 3D. The oldest is Topologically Massive Gravity (TMG) \cite{Deser:1982vy, Deser:1981wh}, in which a gravitational Chern-Simons term gives rise to a single massive graviton of definite helicity. New Massive Gravity, based on a particular curvature-squared invariant, propagates instead a parity doublet of massive helicity-$\pm 2$ modes, and General Massive Gravity combines the two mechanisms \cite{Bergshoeff:2009hq,Bergshoeff:2009aq}. These theories can be tuned to be free of ghosts and tachyons in the bulk, but around AdS vacua this requirement is incompatible with unitarity of the would-be dual CFT, since the two central charges cannot then both be positive.

Minimal Massive Gravity (MMG) \cite{Bergshoeff:2014pca} evades this tension between bulk and boundary unitarity. In a region of its parameter space the bulk graviton is neither a ghost nor a tachyon while both boundary central charges are positive \cite{Arvanitakis:2014xna, Bergshoeff:2019rdb}. This comes with an unusual structure of the dynamics. The MMG field equation differs from that of TMG by a symmetric tensor quadratic in the curvature and does not follow from the variation of any local action for the metric alone, in contrast to TMG, New and General Massive Gravity. The Bianchi identity therefore does not by itself guarantee the consistency of the field equation. Instead, its covariant divergence vanishes only upon a second use of the field equation itself. Theories consistent in this sense are called ``third-way consistent'', and reviewed in \cite{Bergshoeff:2015zga, Deger:2021ojb}. MMG provided the first gravitational example, and the mechanism has since been realized in gauge theories as well, notably in a third-way consistent deformation of 3D Yang-Mills theory \cite{Arvanitakis:2015oga} and higher dimensional p-form theories \cite{Broccoli:2021pvv}.

MMG thus combines a propagating bulk graviton with the prospect of a unitary dual CFT, and it is then natural to ask whether this structure survives coupling to additional fields. Bosonic matter couplings of spin $s< 2$ were constructed in \cite{Arvanitakis:2014yja}, a nontrivial problem since third-way consistency obstructs minimal coupling and requires the stress tensor to enter the field equation through a modified source term. Couplings to fermions were considered in \cite{Cebeci:2020amh}, and a supersymmetric version of MMG was obtained in \cite{Deger:2022gim, Deger:2023eah}. In this paper we consider the coupling of MMG to bosonic higher-spin fields. Apart from its intrinsic interest, this construction provides a setting in which to examine the fate of higher-spin gauge symmetry in the presence of a massive graviton, as well as a starting point for the study of asymptotic symmetries and holography in a massive gravity that may admit a unitary regime.

Since massive gravities propagate local degrees of freedom, they are not Chern-Simons theories. They do, however, admit a ``Chern-Simons-like'' formulation \cite{Hohm:2012vh, Bergshoeff:2014bia}, namely a first-order action for a collection of Lorentz-vector-valued one-forms comprising the dreibein, the (dualized) spin connection and additional auxiliary fields. Higher-spin couplings can then be obtained by using this action  and the strategy of \cite{Campoleoni:2010zq}, promoting the Lorentz-vector-valued fields to $\mathfrak{sl}(N)$-valued ones, as was done for TMG in \cite{Chen:2011vp, Bagchi:2011vr, Bagchi:2011td, Chen:2011yx, Chen:2012ana}. We follow this route for MMG, taking as our starting point the action principle found in \cite{Deger:2022gim}. We further explore a natural extension involving a group-valued scalar $\Phi\in \mathrm{SL}(N)$, which enters as a Stückelberg field for the higher-spin translations. The ordinary formulation is then recovered upon gauge-fixing $\Phi=\mathbb{I}$. The transformation properties of this scalar echo those of the zero-form master field of the Prokushkin–Vasiliev theory \cite{Prokushkin:1998bq}. For related work on the delicate problem of coupling matter to higher-spin theories in three dimensions, see \cite{Prokushkin:1998vn, Chang:2011mz, Kessel:2015kna, Bonezzi:2015igv, Sharapov:2024euk}.

The paper is organized as follows. In Section~\ref{model} we review the first order formulation of MMG. In Section~\ref{hs} we couple it to higher-spin fields up to spin $N$, recovering higher-spin TMG in a suitable limit. We then develop the extended formulation based on the group-valued scalar and show that it carries two independent $\mathfrak{sl}(N)$ gauge symmetries which restore the higher-spin translations. In Section~\ref{solutions} we construct exact solutions, including AdS$_2\times$S$^1$ backgrounds for $N=3$ and $N=4$ in which the higher-spin fields enlarge the radius of the S$^1$. The higher-spin hair supported by these backgrounds requires the MMG deformation and is absent in higher-spin Chern--Simons gravity. In Section~\ref{spectrum} we linearize around the AdS$_3$ vacuum and find that each spin-$j$ field, with $j>2$, gives rise to two massive modes, of spin $j$ and spin $(j-2)$. We conclude in Section~\ref{conclusion}, while conventions and explicit matrix representations are collected in Appendix~\ref{matrix}.

\section{Review of MMG} \label{model}

In this section we review the dynamics of Minimal Massive Gravity in three dimensions \cite{Bergshoeff:2014pca}. We follow \cite{Deger:2022gim, Deger:2023eah} and adopt its first-order formulation, which employs, in addition to the dreibein $e_\mu{}^a$, two independent spin connections $\omega_\mu{}^a$ and $\varpi_\mu{}^a$. The Lagrangian reads\footnote{We set the three-dimensional gravitational constant to $G_3=1$.}%
\be\label{eq:MassiveGravity}
\calL[e,\omega,\varpi] = 
\varepsilon^{\mu\nu\rho}\Big(e_\mu{}^{a}R[{\omega}]_{\nu\rho, a}
+\lambda\, \varepsilon_{a b c} e_\mu{}^{a} e_\nu{}^{b} e_\rho{}^{c}
+ \tau\,e_\mu{}^{a}{} D[\varpi]_\nu  e_{\rho a} \Big)
+\kappa\,{\cal L}_{\rm CS}[\varpi] \, ,
\ee
with $\kappa\tau \neq 0$,\footnote{Note that this formulation of MMG is more general than that of \cite{Bergshoeff:2014pca}, whose parameter space region is restricted to $\kappa\tau<0$.
} and where $\calL_{\rm CS}$ denotes the SO(1,2) Chern-Simons Lagrangian
\be
\calL_{\rm CS}[\varpi] = \epsilon^{\mu\nu\rho}(\varpi_\mu\,^{a}\partial_{\nu}\varpi_{\rho a} + \frac13 \epsilon_{abc} \varpi_\mu\,^a \varpi_\nu\,^b \varpi_\rho\,^c ) \, .
\ee
Here and below, the torsion and curvature of a Lorentz connection $\Omega_\mu{}^a$ are defined as
\begin{align}
T[\Omega]_{\mu\nu}\,^a &:= 2 D[\Omega]_{[\mu} e_{\nu]}\,^a = 2 \partial_{[\mu}e_{\nu]}\,^{a} + 2 \epsilon_{bc}{}^a \Omega_{[\mu}\,^{b} e_{\nu]}\,^{c}\,,\nonumber \\
R[\Omega]_{\mu\nu}\,^a &:=2 \partial_{[\mu}\Omega_{\nu]}\,^{a} + \epsilon_{bc}{}^a \Omega_{\mu}\,^{b} \Omega_{\nu}\,^{c}\,,
\end{align}
with $D[\Omega]_\mu$ the associated Lorentz covariant derivative. These definitions apply to both $\omega$ and $\varpi$. 

Both spin connections can be eliminated on-shell in favour of the dreibein. Varying \eqref{eq:MassiveGravity} with respect to $\omega$ imposes the vanishing of its torsion, $T[\omega]_{\mu\nu}{}^{a}=0$, so that $\omega$ is the Levi-Civita connection of the dreibein. The field equation for the latter is algebraic in $\varpi-\omega$ and gives
\begin{equation} \label{eq:solvarpi}
    \varpi_{\mu}{}^{a} = \omega_\mu{}^a - \frac{1}{\tau} \left(S_{\mu \nu}\, e^{\nu a} + \frac{3 \lambda}{2} e_{\mu}{}^{a} \right) \,,
\end{equation}
where $S_{\mu \nu}= R[\omega]_{\mu \nu}-  \frac{1}{4} R[\omega]\, g_{\mu \nu}$ is the Schouten tensor. Substituting this expression into the field equation for $\varpi$ leads to the metric field equation of MMG \cite{Bergshoeff:2014pca},
\begin{equation}
\left(1+\frac{3\lambda}{2\tau^2}\right)G_{\mu\nu}-\left(\frac{\tau}{\kappa}+\frac{9\lambda^2}{4\tau^2}\right)g_{\mu\nu} -\frac{1}{\tau}C_{\mu\nu} = -\frac{1}{2\tau^2} \epsilon_{\mu\kappa\lambda}  \epsilon_{\nu\sigma\tau} S^{\kappa\sigma} S^{\lambda\tau}\,,
\label{eq:MMGh}
\end{equation}
where $G_{\mu\nu}= R[\omega]_{\mu \nu}-  \frac{1}{2} R[\omega]\, g_{\mu \nu}$ is the Einstein tensor, and $C_{\mu\nu}= \epsilon_{\mu\rho\sigma}\nabla^\rho S^\sigma_{\,\, \nu}$ the Cotton tensor. A key feature of \eqref{eq:MMGh} is that it does not follow from the variation of any local action for the metric alone. Consequently, while the left-hand side of \eqref{eq:MMGh} is divergence-free by virtue of the Bianchi identities, this is not identically the case for the right-hand side, whose divergence vanishes only upon iterated use of \eqref{eq:MMGh} itself, a mechanism known as ``third-way consistency''.\footnote{In contrast, the TMG metric field equation \cite{Deser:1981wh}, which corresponds to \eqref{eq:MMGh} with a vanishing right-hand side, does follow from an action and is consistent by the Bianchi identities alone.}

In this first order formulation of MMG, the Lagrangian \eqref{eq:MassiveGravity} is the sum of the so-called `standard' and the `exotic' actions for three-dimensional gravity \cite{Witten:1988hc},
but with both sectors carrying different spin connections $\omega$ and $\varpi$, respectively. The massive degree of freedom of the model precisely originates in the fact that these connections do not coincide on-shell. Indeed, upon setting $\omega=\varpi$, the Lagrangian \eqref{eq:MassiveGravity} simply reduces to the three-dimensional (topological) gravity of \cite{Mielke:1991nn}, leaving no local degrees of freedom.

We close this section by recalling the maximally symmetric vacua of MMG and the region of parameter space in which it is unitary. Setting $G_{\mu\nu}=-\Lambda\,g_{\mu\nu}$, the Cotton tensor drops out of \eqref{eq:MMGh} and one is left with a quadratic equation for the cosmological constant $\Lambda$,
\begin{equation}
    (\Lambda +3\lambda)^2 + 4\tau^2 \left(\Lambda +\frac{\tau}{\kappa}\right)= 0 \, ,
\label{cosmologicalconstant}
\end{equation}
so that the model admits two (A)dS vacua, with
\begin{equation}
\Lambda_\pm = -\tau^2 \Big(
 (1\pm \Gamma)^2 + \frac{1}{\kappa\tau} \Big)
 \,,
 \label{Lpm}
\end{equation}
where
\begin{equation}
\Gamma = \sqrt{\frac{3\lambda}{\tau^2}-\frac{1}{\kappa\tau}+1}
\;.
\label{eta_Gamma}
\end{equation}
Bulk and boundary unitarity are simultaneously satisfied \cite{Bergshoeff:2014pca, Arvanitakis:2014xna} in the region
\begin{equation}
- \Big(\frac{3\lambda}{2\tau^2}\Big)^2 < \frac{1}{\kappa\tau} < 0
  \,,\;\;\mbox{and}\;\;
\lambda>0
\;,
\label{unitarity}
\end{equation}
where $\Gamma$, and hence both roots $\Lambda_\pm$, are real and the two vacua are AdS, with radii $\ell_{\pm}$ given by $\Lambda_\pm=-1/\ell_\pm^2$. Elsewhere in parameter space the two roots may coincide, $\Lambda_+=\Lambda_-=: \Lambda_m$, which happens when $\Gamma=0$. This locus is known as the \textit{merger point} and corresponds to
\begin{align}
    3\lambda= \frac{\tau}{\kappa} -\tau^2 \,  \implies \, \Lambda_m= - \frac{\tau}{\kappa}(1+\kappa \tau) \, .
\label{merger}
\end{align}

\section{Higher-spin extension} \label{hs}

To couple MMG to higher-spin fields we follow \cite{Campoleoni:2010zq} and promote the $\mathfrak{sl}(2,\bbR)$ algebra-valued fields $\omega_{\mu}\,^a$, $e_\mu\,^a$, and $\varpi_\mu{}^a$ to $\mathfrak{sl}(N,\bbR)$ algebra-valued fields.\footnote{Most of our construction straightforwardly extends to more general algebra $\mathfrak{g}\supset \mathfrak{sl}(2,\mathbb{R})$.} We thus consider 
\begin{equation}
    e_\mu = e_\mu{}^{\cal A} T_{\cal A}\,,\qquad
    \omega_\mu = \omega_\mu{}^{\cal A} T_{\cal A}\,,\qquad
    \varpi_\mu = \varpi_\mu{}^{\cal A} T_{\cal A}\,,\label{eq:fieldsex}
\end{equation}
where the $SL(N,\mathbb{R})$ generators $\{T_{\cal A}\}$ satisfy
\begin{equation}
[T_{\cal A}, T_{\cal B}] = f_{\cal AB}{}^{\cal C}\,T_{\cal C}\,,
\label{eq:algebra}
\end{equation}
with structure constants $f_{\mathcal A\mathcal B}{}^{\mathcal C}$. Note that $f_{\mathcal A\mathcal B\mathcal C}:=f_{\mathcal A\mathcal B}{}^{\mathcal D}\eta_{\mathcal D \mathcal C}$ is fully antisymmetric. Adjoint indices are raised and lowered using the non-degenerate Cartan-Killing form $\eta_{\cal AB}$ over $\mathfrak{sl}(N,\mathbb{R})$. We reserve the use of lower case indices to the adjoint representation of $\mathfrak{sl}(2,\mathbb{R})\cong\mathfrak{so}(1,2)$,
\begin{equation}
    \{T_a\} \subset \{T_{\cal A}\}\,,\qquad a=1,2,3\,.
\end{equation}
The higher spin nature of the fields \eqref{eq:fieldsex} follows from the principal embedding of $\mathfrak{sl}(2,\mathbb{R}) \subset \mathfrak{sl}(N,\mathbb{R})$. Accordingly, the fundamental and the adjoint representations branch into
\begin{align}
    {\bf N} \ \longrightarrow \ & {\bf N}\,,
    \nonumber\\
    \mathfrak{sl}(N,\bbR) \ \longrightarrow \ & 
    {\bf 3} \oplus {\bf 5} \oplus {\bf 7} \oplus \dots \oplus {\bf (2N-1)}
    \,.
    \label{eq:principal}
\end{align}
 For $N=3$, this corresponds to the coupling of a spin-3 field to gravity. For general $N>2$, this is a collection of spin-3, 4, $\dots, N$ fields, coupled to gravity. Further details of this embedding, along with explicit matrix representations for $N=3,4$, are provided in Appendix~\ref{matrix}.

\subsection{Dynamics and symmetries}\label{subsec:dyn}
Given the fields \eqref{eq:fieldsex}, it is convenient to introduce the following definitions for a generic $\mathfrak{sl}(N,\mathbb{R})$-valued connection $\Omega_\mu = \Omega_\mu{}^{\cal A}T_{\cal A}$, in direct parallel with the definitions of Section \ref{model}. Its associated covariant derivative acts on any adjoint-valued field $V^{\cal A}$ as
\begin{equation}
D[\Omega]_\mu V^{\cal A} := \partial_\mu V^{\cal A} + f_{\cal BC}{}^{\cal A}\,\Omega_\mu{}^{\cal B}\,V^{\cal C}\,,
\label{eq:covder}
\end{equation}
while its curvature and its torsion read
\begin{align}
R[\Omega]_{\mu\nu}{}^{\cal A} &:= 2\,\partial_{[\mu}\Omega_{\nu]}{}^{\cal A} + f_{\cal BC}{}^{\cal A}\,\Omega_{\mu}{}^{\cal B}\Omega_{\nu}{}^{\cal C}\,,
\nonumber\\[1ex]
T[\Omega]_{\mu\nu}{}^{\cal A} &:= 2\,D[\Omega]_{[\mu}e_{\nu]}{}^{\cal A} = 2\,\partial_{[\mu}e_{\nu]}{}^{\cal A} + 2\,f_{\cal BC}{}^{\cal A}\,\Omega_{[\mu}{}^{\cal B}e_{\nu]}{}^{\cal C}\,,
\label{eq:tors}
\end{align}
and the associated Chern-Simons Lagrangian is
\begin{equation}
{\cal L}_{\rm CS}[\Omega] := \varepsilon^{\mu\nu\rho}\left(\Omega_\mu{}^{\cal A}\partial_\nu\Omega_{\rho\,{\cal A}} + \frac{1}{3}\,f_{\cal ABC}\,\Omega_\mu{}^{\cal A}\Omega_\nu{}^{\cal B}\Omega_\rho{}^{\cal C}\right)\,.
\label{eq:CS}
\end{equation}
In what follows these definitions are used for both connections $\omega$ and $\varpi$.

As a natural higher-spin generalisation of the dynamics \eqref{eq:MassiveGravity}, we propose the Lagrangian
\begin{equation}
{\cal L}[e,\omega,\varpi] =
\varepsilon^{\mu\nu\rho}\Big(e_\mu{}^{{\cal A}}R[{\omega}]_{\nu\rho, {\cal A}}
+\lambda\, f_{\cal ABC}\,e_\mu{}^{\cal A} e_\nu{}^{\cal B} e_\rho{}^{\cal C}
+ \tau\,e_\mu{}^{\cal A}{} D[\varpi]_\nu  e_{\rho \,{\cal A}} \Big)
+\kappa\,{\cal L}_{\rm CS}[\varpi] 
\,.
\label{eq:LMMG-higher-spin}
\end{equation}
For $N=2$, the $\mathfrak{sl}(N,\mathbb R)$ structure constants reduce to $\epsilon_{abc}$ and \eqref{eq:LMMG-higher-spin} simply reproduces \eqref{eq:MassiveGravity}. Varying with respect to $e^{\cal A}$, $\varpi^{\cal A}$ and $\omega^{\cal A}$ gives, respectively, the field equations
\bea
\label{Eq:EOMMMG2}
\delta e^\mathcal A &:& 
R[{\omega}]_{\mu\nu,{\cal A}}
+ \tau\,T[\varpi]_{\mu\nu,{\cal A}}
+3\,\lambda\,f_{{\cal ABC}}\,e_\mu{}^{\cal B} e_\nu{}^{\cal C} = 0
\;,\nonumber\\[1ex]
\delta {\varpi}^\mathcal A &:& 
R[\varpi]_{\mu\nu,{\cal A}} 
+ \frac{\tau}{\kappa}\,f_{\cal ABC}\,
    e_\mu{}^{\cal B}{}   e_{\nu}{}^{\cal C} = 0 \;,
    \nonumber\\[1ex]
 \delta \omega^\mathcal A &:& 
T[{\omega}]_{\mu\nu,{\cal A}} =0   \;.
    \label{eom_bosonic}
\eea
Let us now comment about the symmetries of the model. The above dynamics is manifestly invariant under local $\mathfrak{sl}(N,\mathbb{R})$ higher-spin Lorentz transformations, acting as
\begin{align}
 \delta e_\mu{}^{\cal A} \ = \ & -f^{\cal A}{}_{\cal BC}\,L^{\cal B}\,e_\mu{}^{\cal C}\,,\nonumber\\  
   \delta \omega_\mu{}^{\cal A} \ = \ & D[\omega]_\mu L^{\cal A}\,,
    \nonumber\\
    \delta \varpi_\mu{}^{\cal A} \ = \ & D[\varpi]_\mu L^{\cal A}\,,
\label{eq:genLor}
\end{align}
with parameter $L^{\cal A}$. For $N=2$ these reduce to the standard $\mathfrak{so}(1,2)$ Lorentz transformations. Each term in \eqref{eq:LMMG-higher-spin} is also invariant, up to a total derivative, under three-dimensional diffeomorphisms, under which all fields transform as one-forms,
\begin{align}
    \delta_\xi e_\mu{}^{\cal A} \ = \ &
    \xi^\nu\partial_\nu e_\mu{}^{\cal A} +
    \partial_\mu \xi^\nu\,e_\nu{}^{\cal A}
    =D[\omega]_\mu\big(\xi^\nu e_\nu{}^{\cal A}\big)-\xi^\nu T[\omega]_{\mu\nu}{}^\mathcal A
    - f^{\cal A}{}_{\cal BC}\,\big(\xi^\nu \omega_\nu{}^{\cal B}\big)\,e_\mu{}^{\cal C}
    \,,
    \nonumber\\
    \delta_\xi \omega_\mu{}^{\cal A} \ = \ &
    \xi^\nu\partial_\nu \omega_\mu{}^{\cal A} +
    \partial_\mu \xi^\nu\,\omega_\nu{}^{\cal A}
    =
    D[\omega]_\mu\big(\xi^\nu\omega_\nu{}^{\cal A}\big)+\xi^\nu R[\omega]_{\nu\mu}{}^{\cal A}
    \,,
    \nonumber\\
    \delta_\xi \varpi_\mu{}^{\cal A} \ = \ &
    \xi^\nu\partial_\nu \varpi_\mu{}^{\cal A} +
    \partial_\mu \xi^\nu\,\varpi_\nu{}^{\cal A} =
    D[\varpi]_\mu\big(\xi^\nu\varpi_\nu{}^{\cal A}\big)+\xi^\nu R[\varpi]_{\nu\mu}{}^{\cal A}
    \,,\label{eq:3Ddiff}
\end{align}
with the diffeomorphism parameter $\xi^\mu$.

In the higher-spin Chern-Simons theory of \cite{Campoleoni:2010zq}, the diffeomorphisms \eqref{eq:3Ddiff} are part of another local $\mathfrak{sl}(N,\mathbb{R})$ symmetry, which we shall refer to as higher-spin translations. In contrast, our model \eqref{eq:LMMG-higher-spin} is not of Chern-Simons type, and the $\tau$ term, which couples the dreibein to $\varpi$ and is responsible for the massive modes, breaks these higher-spin translations down to \eqref{eq:3Ddiff}. In section \ref{hs2} we discuss an extended formulation of the higher-spin MMG model \eqref{eq:LMMG-higher-spin} in which the higher-spin translations are realized through the introduction of a Stückelberg compensating field.

\subsection{TMG limit}
\label{sec:TMG}

Both MMG and TMG propagate a single massive graviton, and the former reduces to the latter in a limit of its parameter space \cite{Bergshoeff:2014pca,Deger:2023eah}. In the first-order formulation this limit extends without modification to the presence of higher-spin couplings. Following \cite{Deger:2023eah}, we set
\begin{equation}
\tau = - \frac{\mu \,(1+\alpha \sigma)^2}{\alpha} \, , \quad
\kappa = \frac{\alpha}{\mu}\, , \quad
\lambda = \frac13\,\alpha\,\Lambda_0-\frac{2\mu^2\,(1+\alpha\sigma)^3}{3\alpha^2}\, ,
\label{rule12}
\end{equation}
and send $\alpha \rightarrow 0$, where $\{\mu,\sigma,\Lambda_0\}$ are the parameters of the MMG formulation of \cite{Bergshoeff:2014pca}. Individually the couplings \eqref{rule12} diverge or vanish in this limit, while their product $\kappa\tau=-(1+\alpha\sigma)^2\to-1$ stays finite, so that the limit is taken within the $\kappa\tau<0$ region. To get a smooth limit, we trade the connection $\varpi_\mu{}^{\cal A}$ for a new field $\beta_\mu{}^{\cal A}$ through
\begin{equation}
\varpi_\mu{}^{\cal A} = \omega_\mu{}^{\cal A} -\frac{3\lambda}{2\,\tau}\,e_\mu{}^{\cal A} + \alpha\,\beta_\mu{}^{\cal A}
\,.
\label{varpiOmega}
\end{equation}
Substituting \eqref{rule12} and \eqref{varpiOmega} into \eqref{eq:LMMG-higher-spin} and expanding in $\alpha$ yields
\bea
{\cal L}&=&
\alpha\,\varepsilon^{\mu\nu\rho}\, \left(
-\sigma\,e_\mu{}^{\cal A} R[\omega]_{\nu\rho,{\cal A}}
+\frac{\Lambda_0}{3}\,f_{{\cal ABC}}\,
    e_\mu{}^{{\cal A}}{} e_\nu{}^{\cal B}  e_{\rho}{}^{\cal C} \right)
+\frac{\alpha}{\mu}\,{\cal L}_{\rm CS}[\omega]
\nonumber\\
&&{}
-\alpha\,\varepsilon^{\mu\nu\rho}
\,\beta_\mu{}^{{\cal A}} T[\omega]_{\nu\rho,{\cal A}}
~+{\cal O}(\alpha^2)
\nonumber\\[2ex]
&=&
\alpha\,{\cal L}_{\rm TMG}[e,\omega,\beta]~+{\cal O}(\alpha^2)
\;.
\label{eq:LTMGlimit}
\eea
At leading order one recovers the first-order Lagrangian of TMG, now $\mathfrak{sl}(N)$-valued, where $\beta_\mu{}^{\cal A}$ appears linearly as a Lagrange multiplier enforcing the torsion constraint $T[\omega]_{\mu\nu}{}^{\cal A}=0$. For $N=3$ this is the TMG of \cite{Chen:2011vp} coupled to a single spin-3 field.

\subsection{Stückelberg formulation} \label{hs2}

In the formulation \eqref{eq:LMMG-higher-spin} the higher-spin translations are broken by the $\tau$ term. As mentioned previously, they can be restored by enlarging the field content with a set of St\"uckelberg scalars. We show below that the resulting model carries two independent local $\mathfrak{sl}(N)$ symmetries, and that gauge-fixing the scalars to a constant returns \eqref{eq:LMMG-higher-spin}.\footnote{The limit described in Section~\ref{sec:TMG} is expected to lead to a similar extended formulation for TMG coupled to higher-spin fields.
} For brevity, we will often use form notation, in which both spacetime and $\mathfrak{sl}(N)$ indices are suppressed and products are understood as matrix products in $\mathfrak{sl}(N)$.

\subsubsection{Extended Lagrangian and symmetries}

We introduce an invertible matrix of scalar fields $\Phi\in\mathrm{SL}(N)$ and consider the Lagrangian
\begin{equation}
{\cal L}[e,\omega,\varpi, \Phi] =
\varepsilon^{\mu\nu\rho}\Big(e_\mu{}^{{\cal A}}R[{\omega}]_{\nu\rho, {\cal A}}
+\lambda\, f_{\cal ABC}\,e_\mu{}^{\cal A} e_\nu{}^{\cal B} e_\rho{}^{\cal C}
+ \frac{\tau}{12\lambda}\,J_\mu{}^{\cal A}{} D[\varpi]_\nu  J_{\rho \,{\cal A}} \Big)
+\kappa\,{\cal L}_{\rm CS}[\varpi] 
\,,
\label{eq:LMMG-higher-spin-J}
\end{equation}
where $\lambda>0$ is assumed. In the $\tau$ term, the dreibein now enters through the scalar current
\begin{equation}
J =  D[\omega]\Phi\, \Phi^{-1}
+\tilde\lambda\,\big(e + \Phi e \Phi^{-1} \big)
= J^{\cal A}\,T_{\cal A}\in\mathfrak{sl}(N)
\,,
\label{eq:current}
\end{equation}
where
\begin{equation}
    D[\omega]\Phi=\partial \Phi+[\omega,\Phi]\,.
\end{equation}
For brevity, we have defined
\begin{equation}
    \tilde\lambda:=\sqrt{3\lambda}\,.
\end{equation}
For $\Phi=\mathbb{I}$ one has $J=2\tilde\lambda\,e$, and \eqref{eq:LMMG-higher-spin-J} reduces to \eqref{eq:LMMG-higher-spin}. This particular choice is discussed in detail in section \ref{sec:gaugefixing}.

We now turn to the two local symmetries of the dynamics \eqref{eq:LMMG-higher-spin-J}. Under the $\mathfrak{sl}(N)$ higher-spin Lorentz transformations \eqref{eq:genLor}, the scalar transforms as
\begin{equation}
    \delta_L \Phi=-[L,\Phi]\,,
\end{equation}
with $L=L^\mathcal A T_{\mathcal A}$. We also introduce the matrix of the adjoint action by
\begin{equation}
    \Phi\, T_{\mathcal A}\,\Phi^{-1}=:\Phi^{\mathcal B}{}_{\mathcal A}\,T_{\mathcal B}\,,
\end{equation}
such that
\begin{equation}
    ( \delta_L \Phi\, \Phi^{-1} )^{\cal A} = 
    (-L + \Phi L \Phi^{-1})^{\cal A}
    = 
    -L^{\cal A} + \Phi^{\cal A}{}_{\cal B}\,L^{\cal B}
    \,.
\end{equation}
It is straightforward to verify that the current \eqref{eq:current} transforms covariantly,
\begin{equation}
    \delta_L J=-[L,J]\,,
\end{equation}
and that each term in the Lagrangian \eqref{eq:LMMG-higher-spin-J} is therefore separately invariant.

The extended formulation admits a second local $\mathfrak{sl}(N)$ symmetry, with parameter $\zeta=\zeta^{\mathcal A}T_{\mathcal A}$, under which the scalars transform as
\begin{equation}
\delta_\zeta \Phi=-\tilde\lambda\,\{\zeta,\Phi\}\,,\qquad
(\delta_\zeta\Phi\,\Phi^{-1})^{\mathcal A}=-\tilde\lambda\,\big(\zeta^{\mathcal A}+\Phi^{\mathcal A}{}_{\mathcal B}\zeta^{\mathcal B}\big)\,,
\end{equation}
while the remaining fields transform as
\begin{align} \label{eq:SL(N)diff}
    \delta_\zeta e_\mu&= \,D[\omega]_\mu\zeta
    \,,\nonumber\\[1ex]
    \delta_\zeta\omega_\mu&=-\tilde \lambda^2\,\big[\zeta,e_\mu\big]\,,\nonumber\\[1ex]
    \delta_\zeta\varpi_\mu&=\tilde\lambda\,D[\varpi]_\mu\zeta
    \,.
\end{align}
These induce the covariant transformations
\begin{equation}
    \delta_\zeta J=-\tilde\lambda\,\big[\zeta,J\big]\,,\qquad
    \delta_\zeta \big(D[\varpi]_\mu J_\nu\big)=-\tilde\lambda\,\big[\zeta,D[\varpi]_\mu J_\nu\big]\,,
\end{equation}
so that \eqref{eq:LMMG-higher-spin-J} is again invariant, now with the sum of the first two terms, the third term, and the last term separately invariant.\footnote{These transformations close on the higher-spin Lorentz symmetry, $[\delta_{\zeta_1},\delta_{\zeta_2}]=\delta_{L_3}$ with $L_3=[\zeta_1,\zeta_2]$.}

\subsubsection{Chern-Simons connections and Stückelberg mechanism}

Both $\mathfrak{sl}(N)$ symmetries can be organised transparently in terms of the Chern-Simons connections
\begin{equation}
    A_\pm=\omega\pm\tilde\lambda\,e\,,
\end{equation}
on which the combined parameters
\begin{equation}
    \epsilon_\pm=L\pm\tilde\lambda\, \zeta\in\mathfrak{sl}(N)
\end{equation}
act as
\begin{equation}
    \delta A_\pm=D[A_\pm]\epsilon_\pm\,.
\end{equation}
The higher-spin Lorentz transformations thus correspond to the diagonal $\mathfrak{sl}(N)$ of $\mathfrak{sl}(N)\oplus\mathfrak{sl}(N)$, and the higher-spin translations the anti-diagonal. In these variables the pair $(A_+,A_-)$ realises the $\mathfrak{sl}(N)\oplus\mathfrak{sl}(N)$ gauge structure of the Chern-Simons theory of \cite{Campoleoni:2010zq}, while $\varpi$ transforms under the $\epsilon_+$ factor alone
\begin{equation}
\delta\varpi=D[\varpi]\epsilon_+\,.
\end{equation}

In this language $\Phi$ is a bi-fundamental field,
\begin{equation}
    \delta\Phi=-\epsilon_+\Phi+\Phi\epsilon_-\,,
\end{equation}
and the current \eqref{eq:current} takes the compact form
\begin{equation}
    J=\big(d\Phi+A_+\Phi-\Phi A_-\big)\Phi^{-1}=:A_+-A_-^\Phi\,,\qquad
    \delta J=-\big[\epsilon_+,J\big]\,,
\end{equation}
where $A_-^\Phi$ denotes the finite gauge transform of $A_-$ by $\Phi$. The current therefore measures the extent to which $A_+$ fails to be the $\Phi$-transform of $A_-$. The scalar $\Phi$ plays the role of a Goldstone, or St\"uckelberg, field for the higher-spin translations and parametrises the symmetric space
\begin{equation}
    \frac{\mathrm{SL}(N)\times \mathrm{SL}(N)}{\mathrm{SL}(N)_\text{diag}}\,,
\end{equation}
where the denominator is the higher-spin Lorentz subgroup.\footnote{The case $N=2$ is of course special in that the massive graviton is then the entire content of the theory. The coset reduces to $\mathrm{SL}(2)\times\mathrm{SL}(2)/\mathrm{SL}(2)_{\rm diag}\cong\mathrm{SL}(2)$, whose three scalars are the compensators that are eaten, and both connections become algebraically solvable, so that one recovers pure MMG with no higher-spin fields.} Its $N^2-1$ components are absorbed into the $\tau$-coupling between $e_\mu$ and $\varpi_\mu$ once the translations are gauge-fixed. We note that the pair $(\Phi,\,J\Phi)$, with $J\Phi=d\Phi+A_+\Phi-\Phi A_-$, is reminiscent of the structure of the zero-form master field and its twisted-adjoint covariant derivative in the Prokushkin-Vasiliev formulation of three-dimensional higher-spin gravity \cite{Prokushkin:1998bq, Bekaert:2004qos}.

\subsubsection{Field equations and gauge fixing}\label{sec:gaugefixing}

Variation of the Lagrangian \eqref{eq:LMMG-higher-spin-J} yields
\bea\label{eom_bosonic_J}
\delta e^{\mathcal A} &:& 
R[{\omega}]_{\mu\nu,{\cal A}}
+ \frac{\tau}{2\tilde\lambda}\,\big(\delta_{\mathcal A}^{\mathcal B}+\Phi^{\mathcal B}{}_{\mathcal A}\big)\, D[\varpi]_{[\mu}J_{\nu]\,\mathcal B}
+\tilde\lambda^2\,f_{{\cal ABC}}\,e_\mu{}^{\cal B} e_\nu{}^{\cal C} = 0
\;,\nonumber\\[1ex]
\delta {\varpi}^{\mathcal A} &:& 
R[\varpi]_{\mu\nu,{\cal A}} 
+ \frac{\tau}{4\kappa\tilde\lambda^2}\,f_{\cal ABC}\,
    J_\mu{}^{\cal B}{}   J_{\nu}{}^{\cal C} = 0 \;,
    \nonumber\\[1ex]
 \delta \omega^{\mathcal A} &:& 
T[{\omega}]_{\mu\nu,{\cal A}}+\frac{\tau}{2\tilde\lambda^2}\big(\delta_{\mathcal A}^{\mathcal B}-\Phi^{\mathcal B}{}_{\mathcal A}\big)\,D[\varpi]_{[\mu}J_{\nu]\,\mathcal B} =0   \,,
\eea
together with the scalar field equation
\begin{equation}
\delta\Phi\,\Phi^{-1}\;:\;   \varepsilon^{\mu\nu\rho}\Big( D[\omega]_{\mu}D[\varpi]_{\nu}J_{\rho}-\big[J_\mu- \tilde\lambda e_\mu,\,D[\varpi]_\nu J_\rho\big]\Big)=0\,,
\label{eq:scalarEOM}
\end{equation}
where we have used the variation of the current,
\begin{align}
\delta_\Phi J &= 
 D[\omega](\delta \Phi\,   \Phi^{-1})
 + [\delta \Phi\,   \Phi^{-1} ,\, D[\omega] \Phi\,   \Phi^{-1} ]
+ \tilde\lambda\,[ \delta \Phi\, \Phi^{-1},\, \Phi e \Phi^{-1} ]\nonumber\\[0.5ex]
&=D[\omega](\delta \Phi\,   \Phi^{-1})+\big[\delta\Phi\,\Phi^{-1},\,J- \tilde\lambda\,e\big]\,.
\end{align}

We may now consider a gauge-fixing of the higher-spin translations. In the gauge $\Phi=\mathbb{I}$, we have
\begin{equation}
    \Phi^{\mathcal A}{}_{\mathcal B}=\delta^{\mathcal A}_{\mathcal B}\,,\qquad J_{\mu}{}^{\mathcal A}=2\,\tilde\lambda\, e_\mu{}^{\mathcal A}\,,\qquad D[\varpi]_{[\mu}J_{\nu]}{}^{\mathcal A}= \tilde\lambda\,T[\varpi]_{\mu\nu}{}^{\mathcal A}\,,
\end{equation}
and the field equations \eqref{eom_bosonic_J} reduce to those of the original formulation \eqref{eom_bosonic}. The higher-spin Lorentz transformations \eqref{eq:genLor} preserve this gauge and remain manifest, whereas the higher-spin translations are broken by the third term of \eqref{eq:LMMG-higher-spin-J}. In this gauge the scalar equation \eqref{eq:scalarEOM} becomes
\begin{equation}
 D[\omega]_{[\mu}T[\varpi]_{\nu\rho]}-\big[e_{[\mu},T[\varpi]_{\nu\rho]}\big]=0\,,
\end{equation}
which is identically satisfied on the shell of \eqref{eom_bosonic} upon using the Bianchi identities
\begin{equation}
    D[\omega]_{[\mu}R[\omega]_{\nu\rho]}=0\,,\qquad D[\omega]_{[\mu} T[\omega]_{\nu\rho]}=\big[R[\omega]_{[\mu\nu},e_{\rho]}\big]\,.
\end{equation}

Finally, it is instructive to discuss the status of 3D diffeomorphisms \eqref{eq:3Ddiff} within this formulation. Off-shell, they can be re-written as a field-dependent $\mathfrak{sl}(N)\oplus\mathfrak{sl}(N)$ transformation with parameters
\begin{equation}
     \zeta=\xi^\mu e_\mu\,,\qquad L=\xi^\mu\omega_\mu\, \qquad\Longleftrightarrow \qquad\epsilon_\pm=\xi^\mu A_{\pm\,\mu}\,,
\end{equation}
supplemented by covariant curvature terms,
\begin{align}
     \delta_\xi \Phi&=\delta_{\zeta,L}\Phi+\xi^\mu J_\mu\Phi\,,\nonumber\\[1ex]
     \delta_\xi e_\mu&=\delta_{\zeta,L}e_\mu+\xi^\nu T[\omega]_{\nu\mu}\,,\nonumber\\[1ex]
     \delta_\xi \omega_\mu&=\delta_{\zeta, L}\omega_\mu+\xi^\nu\big(R[\omega]_{\nu\mu}+\tilde\lambda^2\,[e_\nu,e_\mu]\big)\,,\nonumber\\[1ex]
     \delta_\xi\varpi_\mu&=\delta_{\zeta, L}\varpi_\mu+\xi^\nu R[\varpi]_{\nu\mu}+D[\varpi]_\mu\big(\xi^\nu(\varpi-A_+)_\nu\big)\,.\label{eq:diffOS}
\end{align}
Let us now consider these transformation in the gauge $\Phi=\mathbb{I}$. The inhomogeneous piece in the first equation
\begin{equation}
    \big(\delta_{\zeta, L}\Phi\big)_{|\Phi=\mathbb{I}}=-2\tilde\lambda\, \xi^\mu e_\mu
\end{equation}
is compensated by the current term, so that the gauge is preserved. Imposing in addition the field equations \eqref{eom_bosonic}, the gauge transformation of the dreibein precisely reduces to a standard 3D diffeomorphism,
\begin{equation}
   \delta_\xi e_\mu= \delta_{\zeta,L}e_\mu\qquad \text{On-shell}\,,
\end{equation}
while on the connections the on-shell equivalence is deformed to
\begin{align}
    \delta_\xi\omega_\mu&=\delta_{\zeta,L}\omega_\mu-\tau\,\xi^\nu T[\varpi]_{\nu\mu}\,,\nonumber\\[1ex]
    \delta_\xi\varpi_\mu&=\delta_{\zeta,L}\varpi_\mu+D[\varpi]_\mu\big(\xi^\nu(\varpi-A_+)_\nu\big)-\frac{\tau}{\kappa}\,\xi^\nu\big[e_\nu,e_\mu\big]\,.\label{eq:osdef}
\end{align}
Note that at $\tau=0$, the diffeormophisms and gauge transformations do precisely coincide on-shell on all fields. This corresponds to the pure Chern-Simons case where the $\varpi$ sector decouples, and where the second term in $\delta_\xi\varpi_\mu$ can be absorbed using the independent gauge invariance $\delta_\sigma\varpi=D[\varpi]\sigma$ of $\mathcal L_{\text{CS}}[\varpi]$, where $\sigma$ acts on nothing else.

\section{Exact solutions} \label{solutions}

We now turn to solutions of the higher-spin MMG field equations \eqref{eom_bosonic}.
\subsection{Embedded solutions} \label{inherited}

We first collect solutions that higher-spin MMG inherits from known theories, before turning in Section~\ref{newsolutions} to genuinely new ones. Two classes are immediate.

First, any solution of pure Chern--Simons higher-spin gravity \cite{Campoleoni:2010zq}, such as the higher-spin black holes \cite{Gutperle:2011kf, Ammon:2012wc} and conical defects \cite{Castro:2011iw}, also solves our model. Consider such a solution, with torsion-free $\omega$ and
\begin{equation}
    R[{\omega}]_{\mu\nu}{}^{\cal A}
=\Lambda\,f_{{\cal BC}}{}^{\cal A}\,e_\mu{}^{\cal B} e_\nu{}^{\cal C}
\,.
\end{equation}
Choosing
\begin{equation} \label{eq:solvarpi2}
    \varpi_\mu{}^{\cal A}= \omega_\mu{}^{\cal A}- \frac{1}{2\tau} \left(\Lambda+ 3 \lambda \right) e_\mu{}^{\cal A} \,,
\end{equation}
the first equation of \eqref{eom_bosonic} is satisfied, while the second reduces to the algebraic relation \eqref{cosmologicalconstant} between $\Lambda$ and the parameters of the model. In particular, for any $N$ the model admits the AdS$_3$ vacuum in which all connections lie in $\mathfrak{sl}(2,\mathbb{R})$, with $\varpi$ given by \eqref{eq:solvarpi2} and $\Lambda$ by \eqref{Lpm}. 

Second, spin-2 MMG is known to admit solutions that are not solutions of Einstein gravity \cite{Arvanitakis:2014yja}. Any such solution also solves higher-spin MMG, with all higher-spin fields set to zero. In a basis where the $\mathfrak{sl}(N,\mathbb{R})$ generators are $\mathfrak{sl}(2,\mathbb{R})$ tensors (see Appendix~\ref{matrix}), the gravitational fields occupy the $\mathfrak{sl}(2)$ adjoint sector $\mathbf 3$ of \eqref{eq:principal} and the higher-spin fields the sectors $\mathbf 5, \mathbf 7, \dots$. Since $\mathbf 3$ is a subalgebra, the components of the field equations \eqref{Eq:EOMMMG2} with free index outside $\mathbf 3$ always involve a higher-spin field, and thus vanish when these are switched off, leaving the spin-2 MMG equations.

A specific example of the second class is the AdS$_2\times{\rm S}^1$ solution of spin-2 MMG, which we now describe in detail, as it will serve as the seed for the higher-spin solutions of Section~\ref{newsolutions}. This geometry is not a solution of spin-2 Einstein's gravity nor of TMG, but is an exact solution of MMG \cite{Arvanitakis:2014yja} at the merger point \eqref{merger} of the theory. The metric reads
\begin{equation}\label{eq:ads2xs1metric}
ds^2 = -\frac{1}{\Lambda_m r^2}\left(-dt^2 + dr^2\right) +R_{S^1}^2 \, d\theta^2 \, ,
\end{equation}
where $\Lambda_m<0$ is the merger-point value of the cosmological constant and $\theta\sim\theta+2\pi$ parametrises the compact $S^1$, of arbitrary radius
\begin{align}
    R_{S^1}^2 =: c_1^2 \neq 0 \, .
\end{align}
As we show below, this radius grows once higher-spin fields are switched on. The spacetime is homogeneous \cite{Charyyev:2017uuu}, has vanishing Cotton tensor, and has constant scalar curvature
\begin{equation} \label{Ricciscalar}
    R = 2\Lambda_m \, .
\end{equation}
It appeared as a supersymmetric solution of the off-shell supersymmetric extensions of New and General Massive Gravity \cite{Alkac:2015lma, Deger:2016vrn, Deger:2018kur}.

A metric invariant under the $\mathfrak{sl}(2)$ subalgebra must take the form $g_{\mu\nu}\propto\mathrm{Tr}(e_\mu e_\nu)$, the normalisation remaining to be fixed. Matching \eqref{eq:ads2xs1metric} requires
\begin{equation}
    g_{\mu\nu} = 2\,{\rm Tr}\big(e_\mu e_\nu\big) = 2\, e_{\mu}{}^a e_{\nu}{}^b \, {\rm Tr}(T_aT_b) \,,
    \label{eq:metric2}
\end{equation}
where $\mathrm{Tr}$ denotes the matrix trace of the spin-2 generators $T_a$ of \eqref{Jmatrices}, and the dreibein reads
\begin{align} \label{spin2dreibein}
  e_t{}^1 = e_r{}^2= \frac{1}{r \, \sqrt{-\Lambda_m}} \, , \qquad
  e_{\theta}{}^3 = c_1 \, .
\end{align}
This background admits several continuations by analytic continuation. Wick-rotating the AdS$_2$ time yields a solution with H$^2\times\mathbb{R}$ geometry, where H$^2$ is the Euclidean hyperbolic plane, while for positive cosmological constant one obtains instead dS$_2\times$S$^1$ and S$^2\times\mathbb{R}$.

We anticipate that, more generally, the normalisation of the metric depends on the highest spin $N$ of the solution. It is fixed by requiring the metric to reduce to \eqref{eq:ads2xs1metric} when the higher-spin fields are switched off. For the principal embedding this gives, for $N\geq 2$ \cite{Castro:2011iw},
\begin{equation}\label{metricnormalization}
    g_{\mu\nu} = \frac{12}{N(N^2-1)}\,\mathrm{Tr}\bigl(e_\mu e_\nu\bigr)= \eta_{\mathcal{A}\mathcal{B}}\, e_\mu{}^{\mathcal{A}}e_\nu{}^{\mathcal{B}} \,,
\end{equation}
so that the Cartan--Killing form is
\begin{align}
    \eta_{\mathcal{A}\mathcal{B}}= \frac{12}{N(N^2-1)}\,{\rm Tr}(T_{\mathcal{A}} T_{\mathcal{B}}) \, .
\end{align}

\subsection{$\mathrm{AdS}_2\times \mathrm{S}^1$ solutions with higher-spin hair} \label{newsolutions}

We now switch on higher-spin fields on the AdS$_2\times{\rm S}^1$ background \eqref{eq:ads2xs1metric}, constructing solutions explicitly for $N=3$ and $N=4$. In contrast to the inherited solutions of Section~\ref{inherited}, these carry genuine higher-spin hair and require the massive deformation, solving neither pure higher-spin Chern--Simons gravity nor spin-2 MMG. The extension to higher $N$ is straightforward with the appropriate $\mathfrak{sl}(N,\mathbb{R})$ algebra.

\subsubsection{$N=3$ solution} \label{N3}
We generalise the AdS$_2\times{\rm S}^1$ solution to MMG coupled nontrivially to a spin-3 field.\footnote{AdS$_2\times{\rm S}^1$ solutions were also found in the higher-spin Chern--Simons theory \cite{Gary:2012ms} and further studied in \cite{Afshar:2012nk, Bertin:2012qw}, but their existence there requires a non-principal embedding of $\mathfrak{sl}(2,\mathbb{R})$ \cite{Ammon:2011nk}, whereas we use the principal one.} Under the principal embedding, the adjoint of $\mathfrak{sl}(3,\mathbb{R})$ decomposes as
\begin{equation}
    \mathbf{8} \rightarrow \mathbf{3} \oplus \mathbf{5} \,,
\end{equation}
where $\mathbf{3}$ is generated by the $T_a$ of \eqref{Jmatrices} and $\mathbf{5}$ by the quintet of spin-3 generators $T_{ab}$ of \eqref{iso2}, so that
\begin{align}
     T_{\cal A} = \{T_a, T_{ab}\} \, .
\end{align}
The direct product structure of the metric \eqref{eq:ads2xs1metric} suggests splitting the $\mathfrak{sl}(2,\mathbb{R})$ adjoint index as $a=(\hat{a},3)$, with $\hat a$ and $3$ labelling the AdS$_2$ and $S^1$ directions respectively. The dreibein then decomposes as
\begin{equation}\label{dreibein1}
    e_\mu{}^{\mathcal{A}} = (e_\mu{}^{\hat{a}}, e_\mu{}^3, e_\mu{}^{\hat{a}\hat{b}}, e_\mu{}^{\hat{a}3}, e_\mu{}^{33}) \, ,
\end{equation}
and the spin connections $\omega_\mu{}^{\mathcal{A}}$ and $\varpi_\mu{}^{\mathcal{A}}$ similarly. Guided by the spin-2 solution \eqref{spin2dreibein}, we now make the ansatz for the dreibein components
\begin{equation} \label{dreibein}
e_t{}^1=e_r{}^2= \frac{c_2}{r} \,, \qquad e_t{}^{13}=e_r{}^{23}= \frac{c_4}{r} \,, \qquad e_\theta{}^3=c_1 \,, \qquad e_\theta{}^{33}=c_3 \,, \qquad e_\mu{}^{\hat{a}\hat{b}}=0 \,,
\end{equation}
with the $c_i$ constant, and for the torsion-free spin connection
\begin{align}
     \omega_\mu{}^3=\omega_\mu{}^{33} \,, \quad \omega_\mu{}^{\hat{a}}= \omega_\mu{}^{\hat{a}3} \,, \quad \omega_\mu{}^{\hat{a}\hat{b}}=0 \,.
\end{align}
Solving the last equation of \eqref{eom_bosonic} then implies
\begin{equation} \label{connection}
    \omega_t{}^{3} = \omega_t{}^{33}=\omega_r{}^{3}=\omega_r{}^{33} = -\frac{1}{r} \,,
\end{equation}
while the second equation admits a non-trivial solution only at the merger point \eqref{merger}, where it fixes
\begin{equation} \label{Eq:cconstraintsspin3}
    c_2^2 = -\frac{1}{\Lambda_m} \,, \qquad c_4 = 0 \,.
\end{equation}
A second branch with $c_2 = 0$ and $c_4 \neq 0$ also exists, related to the first by an inner $\mathfrak{sl}(3,\mathbb{R})$ automorphism. We focus on the branch presented above without loss of generality. Finally, the first equation of \eqref{eom_bosonic} fixes the torsionful spin connection $\varpi_\mu{}^{\mathcal{A}}$. Its non-vanishing components are
\begin{align}  \label{barOmega}
    \varpi_t{}^{3} = \varpi_r{}^{3} = -\frac{1}{r} \, , \quad
    \varpi_t{}^{1} = \varpi_r{}^{2} = \frac{\tau}{r\,\sqrt{-\Lambda_m}} \, , \quad
    \varpi_\theta{}^{3} = -\frac{c_1}{\kappa} \, , \quad
    \varpi_\theta{}^{33} = -\frac{c_3}{\kappa} \, .
\end{align}
All field equations of \eqref{eom_bosonic} are now satisfied, with the parameters $c_1$ and $c_3$ left free. With the normalisation \eqref{metricnormalization}, that is
\begin{equation} \label{eq:metric3}
    g_{\mu\nu} = \frac12\,{\rm Tr}\big(e_\mu e_\nu\big) \, ,
\end{equation}
one recovers the metric \eqref{eq:ads2xs1metric} with
\begin{align}
   R_{S^1}^2 = c_1^2 + 3 c_3^2 \, ,\label{eq:radsl3}
\end{align}
and a Ricci scalar matching the spin-2 result \eqref{Ricciscalar}.

Finally, to exhibit the spin-3 hair of the solution in a gauge-invariant way, we compute the field \cite{Campoleoni:2011hg}
\begin{equation}\label{eq:spin3sl3}
    \phi_{\mu\nu\rho}^{(3)} = {\rm Tr}\big(e_{\mu} e_\nu e_{\rho}\big)
    = e_{\mu}{}^{\mathcal{A}}e_{\nu}{}^{\mathcal{B}} e_{\rho}{}^{\mathcal{C}}\, {\rm Tr}\big(T_{\mathcal{A}} T_{\mathcal{B}}T_{\mathcal{C}}\big) \, ,
\end{equation}
which is the spin-3 analogue of the metric. Its only non-vanishing components are
\begin{equation}
    \phi_{tr\theta}^{(3)} = -\frac{2 c_3}{r^2 \Lambda_m} \,,\qquad
    \phi_{\theta\theta\theta}^{(3)} = 6(c_1 - c_3)c_3(c_1 + c_3) \,.
    \label{spin3}
\end{equation}
Since $\phi^{(3)}$ vanishes precisely when $c_3=0$, the enlargement of the radius $R_{S^1}$ in \eqref{eq:radsl3} is genuinely due to the spin-3 hair rather than a spin-2 deformation. This enlargement can also be read as an angle excess. This effect can also be read geometrically. Rescaling $\tilde\theta=(R_{S^1}/c_1)\theta$ removes $c_3$ from the metric but modifies the period of the angular coordinate to $2\pi R_{S^1}/c_1$, an invariant imprint of the spin-3 hair.

\subsubsection{$N=4$ solution}
The construction extends naturally to $\mathfrak{sl}(4,\mathbb{R})$ higher-spin fields coupled to MMG. Under the principal embedding, the adjoint of $\mathfrak{sl}(4,\mathbb{R})$ decomposes as
\begin{equation}
    \mathbf{15} \rightarrow \mathbf{3} \oplus \mathbf{5} \oplus \mathbf{7} \,,
\end{equation}
corresponding to spin-2, spin-3, and spin-4 multiplets, generated by
\begin{align}
     T_{\cal A} =\{T_a, T_{ab}, T_{abc}\} \, .
\end{align}
Explicit representations can be found in Appendix~\ref{matrix}.

Splitting the $\mathfrak{sl}(2,\mathbb{R})$ index as in the $N=3$ case, the dreibein decomposes as
\begin{equation}
    e_\mu{}^{\mathcal{A}} = (e_\mu{}^{\hat{a}}, e_\mu{}^3,   e_\mu{}^{\hat{a} \hat{b}}, e_\mu{}^{\hat{a}3},  e_\mu{}^{33}, 
    e_\mu{}^{\hat{a} \hat{b}\hat{c}},
     e_\mu{}^{\hat{a} \hat{b}3}, e_\mu{}^{\hat{a}33},
     e_\mu{}^{333}
    ) \, ,
  \end{equation}
and the spin connections $\omega_\mu{}^{\mathcal{A}}$ and $\varpi_\mu{}^{\mathcal{A}}$ similarly. In addition to the ansatz \eqref{dreibein}, we take
\begin{align}
    e_t{}^{133}=e_r{}^{233}= \frac{c_6}{r} \, , \quad e_\theta{}^{333} =c_5 \, , \quad e_\mu^{\hat{a}\hat{b}3}=e_\mu^{\hat{a}\hat{b}\hat{c}} =e_\mu^{\hat{a}\hat{b}}= 0 \, ,
\end{align}
and for the spin connection $\omega_\mu{}^{\mathcal{A}}$,
\begin{align}
    \omega_\mu^{333}= \omega_\mu^{33}= \omega_\mu^{3},
\quad \omega_\mu^{\hat{a}33}= \omega_\mu^{\hat{a}3} =\omega_\mu^{\hat{a}},  
\quad \omega_\mu^{\hat{a}\hat{b}3}=\omega_\mu^{\hat{a}\hat{b}\hat{c}}=\omega_\mu^{\hat{a}\hat{b}}=0 \, .
\end{align}
Following the same steps as in the $N=3$ case, the field equations \eqref{eom_bosonic} again select the merger point \eqref{merger} and fix
\begin{equation}
    c_2^2 = -\frac{1}{\Lambda_m} \,, \qquad c_4 = c_6 = 0 \,,
\end{equation}
while the moduli remain free. The torsion-free spin connection $\omega$ retains the form \eqref{connection}, 
\begin{equation} 
    \omega_t{}^{3} = \omega_t{}^{33}= \omega_t{}^{333}=
    \omega_r{}^{3}=\omega_r{}^{33} =\omega_r{}^{333}= -\frac{1}{r} \, ,
\end{equation}
and the torsionful connection $\varpi$ acquires, in addition to the components \eqref{barOmega},
\begin{align}
   \varpi_\theta{}^{333} = -\frac{c_5}{\kappa} \, . 
\end{align}

The higher-spin hair is characterized by the fields \cite{Campoleoni:2011hg, Tan:2011tj}
\begin{align} \label{spin33}
    \Phi^{(3)}_{\mu\nu\rho} &= \frac{1}{4} \mathrm{Tr}\bigl(e_{\mu} e_\nu e_{\rho}\bigr) \,,\nonumber \\
    \Phi^{(4)}_{\mu\nu\rho\sigma} &= \mathrm{Tr}\bigl(e_{\mu} e_\nu e_\rho e_{\sigma}\bigr) 
    - \beta\, \mathrm{Tr}\bigl(e_{\mu} e_\nu\bigr)\,\mathrm{Tr} \bigl(e_\rho e_{\sigma}\bigr) \,.
\end{align}
Requiring the spin-2 solution to be recovered when the higher-spin fields are switched off, that is when $c_3=c_5=0$, fixes $\beta = 41/4$ uniquely.\footnote{This factor was found to be $\beta= 41/10$ in \cite{Campoleoni:2011hg} (see its equation~(3.16)). The mismatch is due to convention differences. In \cite{Campoleoni:2011hg} the $\mathfrak{sl}(2)$ subalgebra generators in $\mathfrak{sl}(4)$ are normalised as $\mathrm{Tr}(T_aT_b)=\eta_{ab}/2$, whereas we have $\mathrm{Tr}(T_aT_b)=5\,\eta_{ab}$ for the $T_a$ of \eqref{Jmatrices}.} The non-vanishing components of these fields are then
\begin{equation}\label{spin3N4}
    \Phi^{(3)}_{tr\theta} = -\frac{ c_3}{r^2 \Lambda_m} \,, \qquad
    \Phi^{(3)}_{\theta\theta\theta} = 3\, c_3 (c_1 - 4 c_5)(c_1 + c_5) \,,
\end{equation}
and 
\begin{equation}\label{spin4N4}
\begin{aligned} 
    \Phi^{(4)}_{tr\theta\theta} &=\frac{4 (2 c_3^2 + 45c_1 c_5 - 5 c_5^2)}{15 r^2 \Lambda_m} \,, \\[6pt]
    \Phi^{(4)}_{\theta\theta\theta\theta} &= -\frac{4}{25} (-85c_1^2  c_3^2 + 16 c_3^4 + 150 c_1^3c_5 + 175c_1^2 c_5^2 - 
   340 c_3^2 c_5^2 - 600 c_1 c_5^3) \, .
\end{aligned}
\end{equation}
Finally, with the normalisation \eqref{metricnormalization}, that is
\begin{equation}\label{metricnormalisation}
    g_{\mu\nu} = \frac15\,\mathrm{Tr}\bigl(e_\mu e_\nu\bigr) \,,
\end{equation}
the metric again takes the form \eqref{eq:ads2xs1metric}, now with
\begin{equation}
    R_{S^1}^2 = c_1^2 + \frac45 c_3^2 + 4 c_5^2\,.
\end{equation}
The scalar curvature remains \eqref{Ricciscalar}, so the AdS$_2 \times S^1$ geometry of the spin-2 sector is preserved by the higher-spin extension, the spin-3 and spin-4 hair modifying only the radius of the circle.

The spin-3 fields \eqref{spin3N4} are switched off by setting $c_3=0$, while the choice $c_5=-c_1$, $c_3=-5c_1$ switches off the spin-4 fields \eqref{spin4N4} and also sets $\Phi^{(3)}_{\theta\theta\theta}=0$. With the spin-4 fields absent, it is natural to ask whether this configuration reduces to a special case of the $N=3$ solution. This requires the physical fields, namely the spin-3 field and the metric, to coincide in the two cases. The spin-3 components \eqref{spin3N4} match those of the $N=3$ solution \eqref{spin3} under the identification $c_3|_{N=4}=2c_3|_{N=3}=:c$, provided the $N=3$ solution is itself specialised to $c_1|_{N=3}=\pm c_3|_{N=3}$, for which $\Phi^{(3)}_{\theta\theta\theta}$ vanishes on both sides. The two solutions then share the same radius $R_{S^1}^2=c^2$. The factor $1/4$ in the $N=4$ spin-3 field \eqref{spin33}, relative to the $N=3$ definition \eqref{eq:spin3sl3}, is what aligns the spin-3 fields under this identification.

\section{Mass spectrum around AdS$_3$} \label{spectrum}

In this section we linearise the higher-spin MMG model around its $\mathrm{AdS}_3$ vacuum and determine the spectrum of fluctuations, identifying the massless higher-spin modes and computing the masses of the massive ones.

\subsection{Linearisation}

In order to linearize the Lagrangian \eqref{eq:LMMG-higher-spin} around the AdS$_3$ vacuum, we split the bosonic fields $e_\mu{}^{\cal A}$, $\omega_\mu{}^{\cal A}$ and $\varpi_\mu{}^{\cal A}$ according to 
\begin{equation} \label{linAnsatz}
  e_{\mu}{}^{{\cal A}} = \mathring{e}_{\mu}{}^{{\cal A}} + \varkappa \, \mathrm{e}_\mu{}^{{\cal A}} \,, \qquad 
  \omega_\mu{}^{{\cal A}} = \mathring{\omega}_\mu{}^{{\cal A}} + \varkappa \, \mathrm{w}_\mu{}^{{\cal A}} \,, \qquad
   \varpi_\mu{}^{{\cal A}} = \mathring{\varpi}_\mu{}^{{\cal A}} + \varkappa\, \mathrm{v}_\mu{}^{{\cal A}} \,,
\end{equation}
where $\mathring{e}_{\mu}{}^{{\cal A}} $ and $\mathring{\omega}_\mu{}^{{\cal A}}$ correspond to the AdS$_3$ geometry with radius $\ell=(-\Lambda)^{-1/2}$ given by (\ref{Lpm}), non-vanishing only on their $\mathfrak{sl}(2,\mathbb{R})$ part. The background field $\mathring{\varpi}_\mu{}^{{\cal A}}$ is given by
\begin{equation} \label{baromega}
   \mathring{\varpi}_\mu{}^{{\cal A}} =\mathring{\omega}_\mu{}^{\cal A} - \frac{1}{2 \tau} (\Lambda + 3 \lambda)\, \mathring{e}_\mu{}^{\cal A}   \,,\
\end{equation}
according to \eqref{eq:solvarpi2}. All fluctuations in (\ref{linAnsatz}) are proportional to $\varkappa$.
Expanding the Lagrangian \eqref{eq:LMMG-higher-spin} with (\ref{linAnsatz}) to 
quadratic order in $\varkappa$, we find
\begin{align} \label{linbosLagr}
 \varkappa^{-2}\, \mathcal{L}_{\rm qu} &= 2\, \varepsilon^{\mu \nu \rho} \,\mathrm{e}_\mu{}^{\cal A} D[\mathring{\omega}]_\nu \mathrm{w}_{\rho \cal A} 
 + \tau \,\varepsilon^{\mu \nu \rho}\, \mathrm{e}_\mu{}^{\cal A} \,D[\mathring{\omega}]_\nu \mathrm{e}_{\rho \cal A} 
 + \kappa \,\varepsilon^{\mu\nu\rho} \,\mathrm{v}_\mu{}^{\cal A} \,D[\mathring{\omega}]_\nu \mathrm{v}_{\rho \,{\cal A}} \nonumber \\
                            & \quad + \varepsilon^{\mu\nu\rho}\, f_{a{\cal BC}} \,\mathring{e}_{\mu}{}^{a}\, \mathrm{w}_\nu{}^{\cal B} \mathrm{w}_\rho{}^{\cal C} 
                            + \frac12 (3 \lambda - \Lambda) \,\varepsilon^{\mu\nu\rho} \, f_{a{\cal BC}} \,\mathring{e}_{\mu}{}^{a}\, \mathrm{e}_\nu{}^{\cal B} \mathrm{e}_\rho{}^{\cal C} \nonumber \\
  & \quad + 2 \tau \, \varepsilon^{\mu\nu\rho}\, f_{a{\cal BC}} \,\mathring{e}_{\mu}{}^{a}\, \mathrm{v}_\nu{}^{\cal B} \mathrm{e}_\rho{}^{\cal C} 
  - \frac{\kappa}{2 \tau}\, (\Lambda + 3 \lambda)\,\varepsilon^{\mu\nu\rho}\, f_{a{\cal BC}} \,\mathring{e}_{\mu}{}^{a}\,\mathrm{v}_\nu{}^{\cal B} \mathrm{v}_\rho{}^{\cal C} \,,
\end{align}
where again lowercase letters correspond to the $\mathfrak{sl}(2,\mathbb{R})$ generators.
At generic points in the parameter space, one can redefine the bosonic fluctuations as
\begin{align}
  \label{eq:diagbosfluct}
  \mathrm{e}_{\mu}{}^{\cal A} &= f^{(-)}_{\mu}{}^{\cal A} + f^{(+)}_{\mu}{}^{\cal A} + p_{\mu}{}^{\cal A}\,, \nonumber \\
  \mathrm{w}_{\mu}{}^{\cal A} &= \frac{1}{\ell}\, f^{(-)}_{\mu}{}^{\cal A} -\frac{1}{\ell} \,f^{(+)}_{\mu}{}^{\cal A}
  -\left(\tau + \frac{1}{2\tau}(\Lambda + 3 \lambda)\right) p_{\mu}{}^{\cal A} \,, \nonumber \\
  \mathrm{v}_{\mu}{}^{\cal A}&= -\frac{1}{\tau} \left( \frac12 \left(3 \lambda + \Lambda\right) - \frac{\tau}{\ell}\right) f^{(-)}_{\mu}{}^{\cal A} 
  -\frac{1}{\tau} \left( \frac12 \left(3 \lambda + \Lambda\right) + \frac{\tau}{\ell}\right) f^{(+)}_{\mu}{}^{\cal A} - \frac{1}{\kappa} \,p_{\mu}{}^{\cal A} \,.
\end{align}
In terms of these new variables, $f^{(\pm)}_{\mu}{}^{\cal A}$, $p_{\mu}{}^{\cal A}$, the Lagrangian (\ref{linbosLagr}) takes  the  diagonalized form
\begin{align} \label{eq:linbosLagrgen}
  \varkappa^{-2}\,  \mathcal{L}_{\mathrm{qu}} &= 
  \beta_+\, \varepsilon^{\mu\nu\rho} f^{(+)}_{\mu}{}^{\cal A} \Big(D[\mathring{\omega}]_{\nu} f^{(+)}_{\rho \,{\cal A}} -\frac{1}{\ell} \, f_{a{\cal AB}} \, f^{(+)}_{\nu}{}^{\cal B} \mathring{e}_\rho{}^{a}  \Big) \nonumber \\ & \ \ \ 
  + \beta_- \, \varepsilon^{\mu\nu\rho} f^{(-)}_{\mu}{}^{\cal A} \Big(D[\mathring{\omega}]_{\nu} f^{(-)}_{\rho\,{\cal A}} + \frac{1}{\ell}   \, f_{a{\cal AB}} \,f^{(-)}_{\nu}{}^{\cal B} \mathring{e}_{\rho}{}^{a} \Big) \nonumber \\ & \ \ \ 
  + \beta_0 \, \varepsilon^{\mu\nu\rho} p_{\mu}{}^{\cal A} \left( D[\mathring{\omega}]_{\nu} p_{\rho \,{\cal A}} + M_p \, f_{a{\cal AB}} \,p_{\nu}{}^{\cal B} \mathring{e}_\rho{}^{a} \right) \,,
 \end{align}
with the constants
\begin{align} \label{ccoeffs}
  \beta_\pm=& \left(\mp \frac{2}{\ell} + \tau + \frac{\kappa}{\tau^2} \left(\frac12 (\Lambda + 3 \lambda) \pm \frac{\tau}{\ell}\right)^2\right) \,, \qquad
  \beta_0 = \left(\frac{1}{\kappa} - \tau - \frac{1}{\tau} (\Lambda + 3 \lambda) \right) \,,
\end{align}
and
\begin{align} \label{eq:Mp}
  M_p=  -\left(\tau + \frac{1}{2\tau}(\Lambda + 3 \lambda) \right) \,.
\end{align}
They satisfy the relation
\begin{equation}
\beta_+\beta_-= -\frac{4\kappa}{\ell^2}\,\beta_0 \, .
\label{alphaalpha}
\end{equation}
It is now straightforward to obtain the field equations for the fluctuations from (\ref{eq:linbosLagrgen}). In the following, we analyze these equations in more detail. We note that the particular components $f_{a{\cal AB}}$ of the structure constants, are actually given by the $\mathfrak{sl}(2,\mathbb{R})$ generators, acting on the full algebra $\mathfrak{sl}(N,\mathbb{R})$
\begin{equation}
  f_{a{\cal AB}}=  -(T_a)_{{\cal AB}} 
\,.
\end{equation}

\subsection{Fluctuation equations}

Let us consider a general fluctuation equation of the type obtained by variation from (\ref{eq:linbosLagrgen})
\begin{equation}
{\cal E}^\mu{}_A =  \varepsilon^{\mu\nu\rho}\,\Big(D[\mathring{\omega}]_{\nu} \phi_{\rho \,{A}} + \mathrm{m} \, (T_a)_{{AB}} \,\phi_{\nu}{}^{B} \mathring{e}_\rho{}^{a} \Big) = 0
 \,,
 \label{eq:generalFluc}
 \end{equation}
with parameter $\mathrm{m}$, around an AdS$_3$ background of radius $\ell$ described by the dreibein field $\mathring{e}_\mu{}^{a}$. Here, we take the field $\phi_{\mu}{}^{{A}}$ to live in some irreducible representation (of a given spin) of $\mathfrak{sl}(2,\mathbb{R})$, and $(T_a)_{{AB}}$ denotes the $\mathfrak{sl}(2,\mathbb{R})$ generator in this representation, \textit{i.e.} we separate the different irreducible parts obtained from variation of (\ref{eq:linbosLagrgen}).

We start by checking the gauge symmetries of this fluctuation equation. With a gauge parameter $\Sigma_A$ and the ansatz
\begin{equation}
\delta_\Sigma \phi_{\mu \,{A}} = D[\mathring{\omega}]_{\mu} \Sigma_{A} 
-\mathrm{m}\,\mathring{e}_{\mu}{}^a (T_a)_{{ AB}} \,\Sigma^{B}
\, ,
\label{eq:gaugep}
\end{equation} 
a variation of (\ref{eq:generalFluc}) yields
\bea
\delta  {\cal E}^\mu{}_A &=&  \frac12\,\varepsilon^{\mu\nu\rho}\, R[\mathring{\omega}]_{\nu\rho}{}^a\,(T_a)_{ AB} \Sigma^{B} 
- \mathrm{m}^2 \, \varepsilon^{\mu\nu\rho}\, \mathring{e}_{\nu}{}^a\mathring{e}_\rho{}^{b}  (T_b)_{{AB}} (T_a)^{{BC}} \, \Sigma_{C} 
\nonumber\\
&=& \frac{1}{2\,\ell^2}\,(1-\mathrm{m}^2\ell^2) \, 
\varepsilon^{\mu\nu\rho}\, \mathring{e}_{\nu}{}^b\mathring{e}_\rho{}^{c} \epsilon_{cab} (T^a)_{{AB}} \, \Sigma^{ B} \, ,
 \eea
where we have used
 \begin{equation}
R[\mathring{\omega}]_{\mu\nu\,a} = -\frac{1}{\ell^2}\,\mathring{e}_\mu{}^{b}\mathring{e}_\nu{}^{c} \, \epsilon_{abc} 
\,,
\label{eq:RinL}
\end{equation}
and the $\mathfrak{sl}(2,\mathbb{R})$ algebra relations for the generators.
This shows that for $\mathrm{m}\ell=\pm1$, equation \eqref{eq:generalFluc} possesses a local gauge symmetry \eqref{eq:gaugep} and accordingly describes a massless higher-spin field. At this critical value the fluctuation equation reduces to the chiral, first-order form of the free higher-spin Chern--Simons theory, its modes being the gauge modes $f^{(\pm)}$ of \eqref{eq:diagbosfluct} that carry no local degrees of freedom.

Let us now be more specific now and take $\phi_\mu{}^{A}$ in the spin-$(j-1)$ representation of $\mathfrak{sl}(2,\mathbb{R})$, with $j>2$. In accordance with the above discussion and embedding (\ref{eq:principal}), such a field will in particular describe a spin $j$ mode. A priori, the full $\mathfrak{sl}(2,\mathbb{R})$ representation content in this field is given by 
\begin{equation}
\phi_\mu{}^{A} \,:\;
[1] \otimes [j-1] = [j-2] \oplus [j-1] \oplus [j]
\,,
\label{eq:spin-content}
\end{equation}
where the AdS$_3$ dreibein $\mathring{e}_\mu{}^a$ has been used to convert spacetime indices into the spin-1 representation of $\mathfrak{sl}(2,\mathbb{R})$.
Contraction of (\ref{eq:generalFluc}) with another derivative and iterating the equation yields
\begin{align}
& \varepsilon^{\mu\nu\rho}\,\Big(D[\mathring{\omega}]_{\mu}  D[\mathring{\omega}]_{\nu} \phi_{\rho \,{A}} -  \mathrm{m}^2 \, (T_a)_{{ AB}} (T_b){}^{B}{}_{C} \,\mathring{e}_\nu{}^{b} \mathring{e}_\rho{}^{a}\, \phi_{\mu}{}^{C}  \Big) = 0
 \,,\nonumber\\
 &\Longrightarrow\qquad  \varepsilon^{\mu\nu\rho}\,\Big(\frac12\,R[\mathring{\omega}]_{\mu\nu}{}^{a}\,(T_a)_{ A}{}^{B} \,\phi_{\rho \,{B}} 
 - \frac12\, \mathrm{m}^2 \, \epsilon_{abc} (T^{a}){}_{A}{}^{B} \,\mathring{e}_\mu{}^{b}\mathring{e}_\nu{}^{c}  \, \phi_{\rho\,B}  \Big) = 0
 \,.
 \end{align}
Using (\ref{eq:RinL}), this implies
\begin{equation}
(1- \mathrm{m}^2\ell^2)\,\mathring{e}^{\mu a}\,(T_a)_{A}{}^{B}\phi_{\mu\,B}  = 0
\,.
\end{equation}
Consequently, unless the field is massless, \textit{i.e.} $\mathrm{m}\ell=\pm1$, the fluctuation equation imposes a non-trivial projection condition
\begin{equation}
\mathring{e}^{\mu a}\,(T_a)_{A}{}^{ B}\phi_{\mu\, B}  = 0
\label{eq:projection}
\,,
\end{equation}
which sets to zero the $[j-1]$ part in the expansion (\ref{eq:spin-content}). For the rest, we will restrict to this massive case.

In order to analyze the field equations for the remaining components of (\ref{eq:spin-content}) let us introduce the
$\mathfrak{sl}(2,\mathbb{R})$ Casimir 
\begin{equation}
{\cal C}_{A}{}^{B} = \tfrac12\,(T_a)_{A}{}^{C}(T^a)_{C}{}^{B}
\,,
\label{eq:Cas1}
\end{equation}
whose eigenvalue (in the spin-$(j-1)$ representation) is given by $\frac12\,j\,(j-1)$.
Similarly, the Casimir operator acting on $\phi_{\mu A}$ then is given by
\begin{equation}
{\cal C}_{\mu A}{}^{\nu B} = \tfrac12\,
\Big((T_a)_{A}{}^{C}\,\delta_\mu^\rho
+\delta_A^C\,\epsilon_{a\mu}{}^{\rho}\Big)
\Big((T^a)_{C}{}^{B}\,\delta_\rho^\nu
+\delta_C^B\,\epsilon^a{}_{\rho}{}^{\nu}\Big)
\,,
\label{eq:Cas2}
\end{equation}
and has eigenvalues $\frac12\,j\,(j+1)$ and $\frac12\,(j-1)(j-2)$, depending on which of the irreducible components in (\ref{eq:spin-content}) it acts on. Comparing the action of (\ref{eq:Cas1}) and (\ref{eq:Cas2}), and their respective eigenvalues, we 
find that
\begin{equation}
\epsilon_{a\mu\nu}\,(T^a)_{{ AB}}\,\phi^{\nu\,{ B}} =
 \left\{
 \begin{array}{ll}
  (1-j) \, \phi^{\mu}{}_{A}
 & \mbox{spin part } [j] \\
 j\,  \phi^{\mu}{}_{A}
 & \mbox{spin part } [j-2] \\
 \end{array}
 \right.
 ,
 \label{eq:CasMix}
  \end{equation}
for the two remaining irreducible representations in (\ref{eq:spin-content}). As a consequence, the fluctuation equation (\ref{eq:generalFluc}) implies
\begin{equation}
 \varepsilon^{\mu\nu\rho}\, D[\mathring{\omega}]_{\nu} \phi_{\rho \,{A}} =
 \left\{
 \begin{array}{ll}
   \mathrm{m}\,(j-1) \, \phi^{\mu}{}_{A}
 & \mbox{spin part } [j] \\
  -   \mathrm{m} \,j\,  \phi^{\mu}{}_{A}
 & \mbox{spin part } [j-2] \\
 \end{array}
 \right.
 ,
 \label{eq:fluc_simp}
  \end{equation}
for the non-vanishing components of the massive field.
We may explicitly parametrize the two irreducible parts within $\phi_{\mu{A}}$ as
\begin{equation}
\phi_{\mu{A}}= \tilde\varphi_{\mu{A}} +  I_{{\mu\,{ A}}}{}^{\nu\,\underline{ B}}\,\varphi_{\nu\,\underline{B}} 
\,,
\end{equation}
with a unique $\mathfrak{sl}(2,\mathbb{R})$ invariant intertwiner $I_{{\mu\,{ A}}}{}^{\nu\,\underline{B}}\,$ and indices $\underline{A}$ labelling the spin-$(j-3)$ representation. 
From the representation content of its indices, we find that
\begin{equation}
I_{{\mu\,{A}}}{}^{\mu\,\underline{B}} = 0 = I_{{[\mu\,|{A}|,}}{}_{\nu]}{}^{\underline{B}}
\,,
\end{equation}
since there is no singlet nor a spin-1 representation in the tensor product ${A} \otimes \underline{B}$. 
The second equation of (\ref{eq:fluc_simp}) then yields
\begin{equation}
   -   \mathrm{m} \,j\,I_{{\mu\,{A}}}{}^{\nu\,\underline{B}}\,\varphi_{\nu\,\underline{B}} 
=
\varepsilon_\mu{}^{\nu\rho}\, I_{{\rho\,{ A}}}{}^{\sigma\,\underline{B}}\, D[\mathring{\omega}]_{\nu} \varphi_{\sigma\,\underline{B}}  
=
\varepsilon_\mu{}^{\nu\rho}\, I_{{\rho\,{ A}}}{}^{\sigma\,\underline{B}}\, D[\mathring{\omega}]_{[\nu} \varphi_{\sigma]\,\underline{B}}  
\,,
\end{equation}
since its projection onto the $[j-2]$ representation can only contain the curl $D[\mathring{\omega}]_{[\nu} \varphi_{\sigma]\,\underline{B}}$.
Upon using the Schouten identity in the three-dimensional spacetime indices, we finally obtain
\begin{equation}
\frac12\,
\varepsilon_{\sigma}{}^{\nu\rho}\, I_{{\mu\,{A}}}{}^{\sigma\,\underline{B}}\, D[\mathring{\omega}]_{\nu} \varphi_{\rho\,\underline{B}} 
=      \mathrm{m} \,j\,I_{{\mu\,{ A}}}{}^{\nu\,\underline{ B}}\,\varphi_{\nu\,\underline{B}} 
\,.
\end{equation}

To summarize, we have thus shown that in the massive case, $\mathrm{m}^2\ell^2\not=1$, the linear equation (\ref{eq:generalFluc}) implies the equations
\begin{align}
 \varepsilon^{\mu\nu\rho}\, D[\mathring{\omega}]_{\nu} \tilde\varphi_{\rho \,{A}} = &\,
   (j-1) \,\mathrm{m}\, \tilde\varphi^{\mu}{}_{A}
   \, , \nonumber\\
 \varepsilon^{\mu\nu\rho}\, D[\mathring{\omega}]_{\nu} \varphi_{\rho \,\underline{{A}}} = &\,
   2\,j\,\mathrm{m} \, \varphi^{\mu}{}_{\underline{{A}}}
\,,
 \label{eq:fluc_simp2}
\end{align}
for the two non-vanishing components of spin $j$ and spin $(j-2)$, respectively. This means these modes carry conformal dimension
\bea
\Delta_{\tilde\varphi} = 1+ (j-1) \,\mathrm{m}\,,\qquad
\Delta_\varphi = 1+ 2j \,\mathrm{m}\,,
\label{eq:Deltas}
\eea
respectively, of the same chirality. Explicitly
\bea
&&
\left(\Delta_{\tilde\varphi,{\rm L}},\Delta_{\tilde\varphi,{\rm R}}\right) = \Big(
\tfrac12(j+1)+ \tfrac12\,(j-1)  \,\mathrm{m},
\tfrac12(1-j)+ \tfrac12\,(j-1) \,\mathrm{m}\Big)
\,,
\nonumber\\
&&
\left(\Delta_{\varphi,{\rm L}},\Delta_{\varphi,{\rm R}}\right) 
= \Big( \tfrac12(j-1)+  j \,\mathrm{m}, \tfrac12(3-j)+ j \,\mathrm{m}\Big)\,.
\label{eq:DeltasLR}
\eea
For spin-3 and the TMG equations, this reproduces the result of \cite{Bagchi:2011vr}
with $\mathrm{m}=\mu\ell$.
For higher spins this may be compared to the results of \cite{Bagchi:2011td,Chen:2011yx} in the TMG case.
For the present model describing higher-spin fields coupled to MMG, the linearized Lagrangian (\ref{eq:linbosLagrgen}) found above, thus 
 yields massive modes (\ref{eq:DeltasLR}) with 
 \begin{align} \label{eq:alphaMMG}
  \mathrm{m} = M_p=  -\left(\tau + \frac{1}{2\tau}(\Lambda + 3 \lambda) \right) \,.
\end{align}

\section{Conclusions and Outlook} \label{conclusion}
In this paper we have considered the coupling of MMG to bosonic fields of spin greater than two. This was done by adapting the method used for three-dimensional Einstein gravity \cite{Campoleoni:2010zq} to the first-order formulation of MMG \cite{Deger:2022gim, Deger:2023eah}. Several interesting directions deserve further exploration.

Among the most natural directions is the asymptotic symmetry analysis, which would characterise the putative dual CFT. For spin-2 MMG the asymptotic algebra is two copies of the Virasoro algebra \cite{Bergshoeff:2014pca}. At a generic point of parameter space we expect these to be enhanced, by analogy with Einstein gravity \cite{Henneaux:2010xg, Campoleoni:2010zq}, to two copies of the classical $\mathcal{W}_N$ algebra with the same central charges. The analysis becomes more delicate at the chiral points, where one central charge vanishes and logarithmic modes appear, as studied for higher-spin TMG in \cite{Chen:2011vp, Bagchi:2011vr} (see also \cite{Bergshoeff:2009tb}). For spin-2 MMG, however, these points lie outside the unitary region. Whether the higher-spin fields preserve unitarity remains an open question. We have worked throughout with the principal embedding of $\mathfrak{sl}(2)$, and other embeddings, with their different spin content, may display a different unitarity structure \cite{Castro:2012bc}.

A metric-like formulation, along the lines of \cite{Campoleoni:2012hp}, would also be valuable. In a third-way consistent theory coupled to matter the divergence of the energy-momentum tensor does not vanish on the matter field equations, but instead obeys the modified condition of \cite{Arvanitakis:2014yja}. Such a formulation would allow this to be verified explicitly for our higher-spin fields.

It would also be interesting to clarify the relation to the unfolded approach to matter-coupled higher-spin gravity of Prokushkin and Vasiliev \cite{Prokushkin:1998bq, Vasiliev:1992gr} (see \cite{Kessel:2016hld} for a review and \cite{Sharapov:2024euk} for a recent discussion). The group-valued scalar of our extended formulation transforms in a way reminiscent of the Prokushkin and Vasiliev zero-form master field. Making this correspondence precise, in particular identifying the finite-$N$ analogue of the twisted-adjoint representation, could perhaps connect the two constructions. The structure and consistent truncation of that representation were analysed in \cite{Kessel:2015kna}. In a related vein, these unfolded methods were used to study linearised topologically massive higher-spin fields in \cite{Boulanger:2014vya}, where discrepancies with the higher-spin TMG proposals of \cite{Bagchi:2011td, Chen:2011yx} were noted. It would be interesting to carry out a similar comparison for higher-spin MMG.

Our extended formulation suggests a further, more speculative direction. The scalar equips the theory with a full $\mathrm{SL}(N)\times\mathrm{SL}(N)$ gauge symmetry, and it is natural to ask whether, upon imposing asymptotically AdS boundary conditions of Drinfeld and Sokolov type \cite{Campoleoni:2010zq, Henneaux:2010xg}, the boundary dynamics organises into an $\mathfrak{sl}(N)$ Toda theory, much as three-dimensional Einstein gravity reduces to Liouville theory \cite{Coussaert:1995zp}. In this picture the scalar could provide a bulk counterpart of the Toda field and the broken higher-spin translations a bulk realisation of its $\mathcal{W}_N$ symmetry. We leave the investigation of this reduction to future work.

On the solutions side, the AdS$_2\times$S$^1$ backgrounds we constructed, which are not solutions of higher-spin Einstein gravity in the principal embedding, invite a study of their holographic interpretation \cite{Gary:2012ms, Afshar:2012nk, Bertin:2012qw}. We expect a broader class of such solutions to exist. Higher-spin black holes would be of particular interest. For spin-2 MMG, AdS$_2\times$S$^1$ arises as the near-horizon geometry of the extremal AdS black hole \cite{Arvanitakis:2014yja}, so its higher-spin generalisation is a natural target. Warped AdS, likewise a solution of spin-2 MMG but not of Einstein gravity \cite{Arvanitakis:2014yja}, is another candidate. Its higher-spin version could be relevant to extensions of the warped AdS/warped CFT correspondence \cite{Anninos:2008fx, Chen:2012ana}. More challenging still would be to find a solution that exists only when the higher-spin fields are active, admitting no consistent truncation to a purely gravitational configuration.

Finally, a supersymmetric extension of our model would be desirable. Since third-way consistent equations do not follow from an action in the usual way, the existence of a supersymmetric completion is not guaranteed, yet one was constructed for spin-2 MMG in \cite{Deger:2022gim, Deger:2023eah}. A natural approach to the higher-spin case would be to replace $\mathfrak{sl}(N)$ by a superalgebra, along the lines by which three-dimensional supergravity is formulated as a Chern--Simons theory of a supergroup \cite{Achucarro:1987vz} and extended to higher spins \cite{Henneaux:2012ny}. 

\section*{Acknowledgements}

We thank Stefan Fredenhagen, Dmitri Sorokin and Stefan Theisen for useful discussions. 
This work has been partially supported by a PHC BOSPHORE, project No 50765SK and by the Scientific
and Technological Research Council of T\"urkiye (T\"ubitak) project 123N953. NSD wishes to thank the Albert Einstein Institute, Potsdam and the ENS de Lyon for hospitality during the course of this work. NSD also wishes to thank the II. Institute for Theoretical Physics at Hamburg University for hospitality during the final phase of this paper and acknowledges support by the DFG – SFB 1624 – “Higher structures, moduli spaces and integrability” –
506632645. The research of FC was funded by the DFG (German Research Foundation) – Project number: 521509185. 

\appendix 
\section{$\mathfrak{sl}(N, \mathbb{R})$ representations and principal embedding} \label{matrix}
In this appendix we give our conventions and explicit representations of the SL($N$) generators, for $N\leq 4$, which were used for the construction of solutions in Section \ref{solutions}. These generators realize the principal embedding of
$\mathfrak{sl}(2)$ into $\mathfrak{sl}(N)$, \textit{i.e.} under the adjoint action of
$\{H,E,F\}$, the $\mathfrak{sl}(N)$ algebra decomposes as
$\mathfrak{sl}(N)\simeq\bigoplus_{s=2}^{N}\mathsf{g}^{(s)}$, where
$\mathsf{g}^{(s)}$ denotes the $(2s-1)$-dimensional spin-$(s{-}1)$ representation of $\mathfrak{sl}(2)$. The generators $W$ and $U$ furnish the $s>2$ mulitplets: $W_{-2},\dots,W_2$ span the $\mathsf{g}^{(3)}$ quintet and, for $\mathfrak{sl}(4)$,
$U_{-3},\dots,U_{3}$ span the $\mathsf{g}^{(4)}$ septet.

\

\noindent $\bullet$ For $\mathfrak{sl}(2, \mathbb{R})$: A standard basis is given by 
\be
\begin{split}
&H = \begin{pmatrix}
\frac12 & 0  \\
0 & -\frac12  
\end{pmatrix} \quad
E = \begin{pmatrix}
0 & 0  \\
1 & 0  
\end{pmatrix} \quad
F = \begin{pmatrix}
0 & -1  \\
0 & 0  
\end{pmatrix}
\end{split}
\ee 
in which the algebra reads
\begin{equation}
[H,E]=-E\,, \quad [H,F]=F\,,\quad[E,F]=2H \, .
\end{equation}
Another convenient basis is given in terms of
\begin{align} \label{Jmatrices}
  T_1=\frac{E+F}{2}\, , \quad T_2=\frac{E-F}{2} \, , \quad T_3=H \, ,
\end{align}
which satisfy
\begin{align}\label{Jalgebra}
    [T_a, T_b]=\epsilon_{abc}T^c \, ,
\end{align}
with $\epsilon_{123}=-1$ and where and indices are raised and lowered with $\eta_{ab}=(-1, 1, 1)$.

\

\noindent $\bullet$ For $\mathfrak{sl}(3, \mathbb{R})$: The standard realization given in \cite{Campoleoni:2010zq} reads
\be
\renewcommand{\arraystretch}{1.1}
\begin{array}{c@{\qquad}c@{\qquad}c}
H = \begin{pmatrix} 1 & 0 & 0 \\ 0 & 0 & 0 \\ 0 & 0 & -1 \end{pmatrix} &
E = \begin{pmatrix} 0 & 0 & 0 \\ 1 & 0 & 0 \\ 0 & 1 & 0 \end{pmatrix} &
F = \begin{pmatrix} 0 & -2 & 0 \\ 0 & 0 & -2 \\ 0 & 0 & 0 \end{pmatrix} \\
\vphantom{\begin{pmatrix}0\\0\\0\\0\end{pmatrix}}
W_0 = \begin{pmatrix} 1 & 0 & 0 \\ 0 & -2 & 0 \\ 0 & 0 & 1 \end{pmatrix} &
W_{1} = \tfrac{3}{2}\begin{pmatrix} 0 & 0 & 0 \\ 1 & 0 & 0 \\ 0 & -1 & 0 \end{pmatrix} &
W_{-1} = \begin{pmatrix} 0 & -3 & 0 \\ 0 & 0 & 3 \\ 0 & 0 & 0 \end{pmatrix} \\[2.5ex]
W_2 = \begin{pmatrix} 0 & 0 & 0 \\ 0 & 0 & 0 \\ 3 & 0 & 0 \end{pmatrix} &
W_{-2} = \begin{pmatrix} 0 & 0 & 12 \\ 0 & 0 & 0 \\ 0 & 0 & 0 \end{pmatrix} 
\end{array}
\ee
For the construction of solutions in Section~\ref{solutions}, it is convenient to work in a different basis, given by the generators $\{T_a, T_{ab}\}$, where the $T_a$ are defined as in \eqref{Jmatrices} from the $3\times3$ matrices $(H,E,F)$. The spin-3 generators $T_{ab}$ are $\mathfrak{sl}(2,\mathbb{R})$ tensors, and as such are symmetric ($T_{ab}=T_{ba}$) and traceless ($T^a{}_a=0$) in their $\mathfrak{sl}(2,\mathbb{R})$ adjoint indices. Note that this does not mean the $T_{ab}$ are symmetric as matrices. A representation in terms of the $W$ matrices above is given in \cite{Campoleoni:2010zq},
\begin{alignat}{3}
& T_{11} \, = \, \frac{1}{4} \left( W_2 + W_{-2} + 2\, W_0 \right) \, , \qquad\qquad & & T_{12} \, = \,  \frac{1}{4} \left( W_2 - W_{-2} \right) \, , \nonumber \\
& T_{22} \, = \, \frac{1}{4} \left( W_2 + W_{-2} - 2\, W_0 \right) \, , \qquad\qquad & & T_{13} \, = \,  \frac{1}{2} \left( W_1 + W_{-1} \right) \, , \nonumber \\
& T_{33} \, = \, W_0 \, , & & T_{23} \, = \,  \frac{1}{2} \left( W_1 - W_{-1} \right) \, , \label{iso2}
\end{alignat}
and they satisfy the algebra
\begin{align}\label{JTalgebra}
 [T_a,   T_{bc}] = 2 \, \epsilon^d{}_{a(b} T_{c)d} \, .
\end{align}
Note that, compared to \cite{Campoleoni:2010zq}, we include a normalisation factor in the symmetrisation and antisymmetrisation, \textit{e.g.}
\begin{equation}
V_{(ab)} = \frac12\,(V_{ab}+V_{ba})
\,.
\end{equation}
The non-vanishing structure constants $f_{\cal A\cal B\cal C}=f_{[\cal A\cal B\cal C]}$ of the algebra spanned by $T_{\cal A}=\{T_a,T_{ab}\}$ can be read off as
\begin{equation} \label{structure}
f_{a,b,c} = \epsilon_{abc}\;,\qquad
f_{a,bc,de} = \epsilon_{ab(d}\eta_{e)c} 
+ \epsilon_{ac(d}\eta_{e)b} 
\,,
\end{equation}
while the Cartan--Killing form is defined as \cite{Campoleoni:2012hp}
\begin{align}
    \eta_{\cal AB}= \frac{1}{2} \mathrm{Tr}(T_{\cal A} T_{\cal B}) \, ,
\end{align}
and equals $\eta_{ab}$ when ${\cal A}=a$, ${\cal B}= b$.

\ 

\noindent $\bullet$ For $\mathfrak{sl}(4,\mathbb{R})$: 
\begin{gather}
H = \tfrac12\begin{pmatrix} -3 & 0 & 0 & 0 \\ 0 & -1 & 0 & 0 \\ 0 & 0 & 1 & 0 \\ 0 & 0 & 0 & 3 \end{pmatrix} \qquad
E = \begin{pmatrix} 0 & 1 & 0 & 0 \\ 0 & 0 & 1 & 0 \\ 0 & 0 & 0 & 1 \\ 0 & 0 & 0 & 0 \end{pmatrix} \qquad
F = \begin{pmatrix} 0 & 0 & 0 & 0 \\ -3 & 0 & 0 & 0 \\ 0 & -4 & 0 & 0 \\ 0 & 0 & -3 & 0 \end{pmatrix}\nonumber
\\[2.5ex]
W_0 = \begin{pmatrix} 1 & 0 & 0 & 0 \\ 0 & -1 & 0 & 0 \\ 0 & 0 & -1 & 0 \\ 0 & 0 & 0 & 1 \end{pmatrix} \qquad
W_1 = \begin{pmatrix} 0 & -1 & 0 & 0 \\ 0 & 0 & 0 & 0 \\ 0 & 0 & 0 & 1 \\ 0 & 0 & 0 & 0 \end{pmatrix} \qquad
W_{-1} = \begin{pmatrix} 0 & 0 & 0 & 0 \\ 3 & 0 & 0 & 0 \\ 0 & 0 & 0 & 0 \\ 0 & 0 & -3 & 0 \end{pmatrix}\nonumber
\\[2.5ex]
W_2 = \begin{pmatrix} 0 & 0 & 1 & 0 \\ 0 & 0 & 0 & 1 \\ 0 & 0 & 0 & 0 \\ 0 & 0 & 0 & 0 \end{pmatrix} \qquad
W_{-2} = \begin{pmatrix} 0 & 0 & 0 & 0 \\ 0 & 0 & 0 & 0 \\ 12 & 0 & 0 & 0 \\ 0 & 12 & 0 & 0 \end{pmatrix} \qquad
U_0 = \tfrac{3}{10}\begin{pmatrix} -1 & 0 & 0 & 0 \\ 0 & 3 & 0 & 0 \\ 0 & 0 & -3 & 0 \\ 0 & 0 & 0 & 1 \end{pmatrix}\nonumber
\\[2.5ex]
U_1 = \tfrac15\begin{pmatrix} 0 & 2 & 0 & 0 \\ 0 & 0 & -3 & 0 \\ 0 & 0 & 0 & 2 \\ 0 & 0 & 0 & 0 \end{pmatrix} \qquad
U_{-1} = \tfrac65\begin{pmatrix} 0 & 0 & 0 & 0 \\ -1 & 0 & 0 & 0 \\ 0 & 2 & 0 & 0 \\ 0 & 0 & -1 & 0 \end{pmatrix} \qquad
U_2 = \tfrac12\begin{pmatrix} 0 & 0 & -1 & 0 \\ 0 & 0 & 0 & 1 \\ 0 & 0 & 0 & 0 \\ 0 & 0 & 0 & 0 \end{pmatrix}\nonumber
\\[2.5ex]
U_{-2} = \begin{pmatrix} 0 & 0 & 0 & 0 \\ 0 & 0 & 0 & 0 \\ -6 & 0 & 0 & 0 \\ 0 & 6 & 0 & 0 \end{pmatrix} \qquad
U_3 = \begin{pmatrix} 0 & 0 & 0 & 1 \\ 0 & 0 & 0 & 0 \\ 0 & 0 & 0 & 0 \\ 0 & 0 & 0 & 0 \end{pmatrix} \qquad
U_{-3} = \begin{pmatrix} 0 & 0 & 0 & 0 \\ 0 & 0 & 0 & 0 \\ 0 & 0 & 0 & 0 \\ -36 & 0 & 0 & 0 \end{pmatrix}
\end{gather}
As before, it is again convenient to work in a basis given by the generators $\{T_a, T_{ab}, T_{abc}\}$, with the $T_a$ defined as in \eqref{Jmatrices}. The generators $T_{ab}$ and $T_{abc}$ are symmetric and traceless in their $\mathfrak{sl}(2,\mathbb{R})$ indices. The $T_{ab}$ are given in \eqref{iso2}, while the explicit form of the $T_{abc}$ in terms of the $U$ matrices was obtained in \cite{Tan:2011tj},
\begin{align}
&T_{333}= U_0,\,\,T_{331}=\frac{1}{2} \left( U_1 + U_{-1} \right),\,\,T_{332} = \frac{1}{2} \left( U_1 - U_{-1} \right),\,\,T_{311}=\frac{1}{4} \left( U_2 + U_{-2} \right) + \frac{1}{2} U_0, \nonumber \\
&T_{123} = \frac{1}{4} \left( U_2 - U_{-2} \right),\,\,T_{322}=\frac{1}{4} \left( U_2 + U_{-2} \right)-\frac{1}{2}U_0, \, T_{111}=\frac{1}{8} \left( U_3 + U_{-3} + 3(U_1+U_{-1}) \right),\nonumber \\
&T_{112}=\frac{1}{8} \left( U_3 - U_{-3} + U_1 - U_{-1} \right),\,\,T_{122}=\frac{1}{8} \left( U_3 + U_{-3} - U_1 - U_{-1} \right),\nonumber \\
&T_{222}=\frac{1}{8} \left( U_3 - U_{-3} + 3(U_{-1} - U_1) \right) \, .
\end{align}
In addition to the commutators \eqref{Jalgebra} and \eqref{JTalgebra}, the algebra now involves
\begin{align}
 [T_a,   T_{bcd}] = 6 \, \epsilon^d{}_{a(b} T_{cd)d} \, ,
\end{align}
which yields the additional non-vanishing structure constant
\begin{align}
    f_{a,bcd,efg}=  \epsilon_{ab(f}\eta_{e|c|}\eta_{g)d} + \epsilon_{ad(f}\eta_{e|b|}\eta_{g)c} + \epsilon_{ac(f}\eta_{e|d|}\eta_{g)b} + \textrm{3 more permutations} \, ,
\end{align}
where the last three terms symmetrise the indices $\{b,c,d\}$. In this case, the Cartan--Killing form is defined as
\begin{align}
    \eta_{\cal AB}= \frac{1}{5} \mathrm{Tr}(T_{\cal A} T_{\cal B}) \, ,
\end{align}
such that again it equals $\eta_{ab}$ when ${\cal A}=a$ and ${\cal B}=b$.

\bibliographystyle{utphys} 
\bibliography{refs2.bib}

\end{document}